\documentclass[review, 5p]{elsarticle}

\usepackage{times,amsmath,color,amssymb,latexsym,graphicx,epic,eepic,verbatim,stmaryrd,algorithmic}
\usepackage[numbers]{natbib}

\journal{Neural Networks}

\biboptions{authoryear}

\begin{document}

\begin{frontmatter}

\title{Biomimetic Race Model of the Loop between the Superior Colliculus and the \\Basal Ganglia: Subcortical Selection of Saccade Targets}

\author[Sorba,ISIR,correspondingauthor]{Charles Thurat\corref{mycorrespondingauthor}}
\cortext[correspondingauthor]{Corresponding author. Tel.: +33 (0)1 44 27 63 81}
\ead{thurat@isir.upmc.fr}
\author[Sorba,ISIR,Sorbb,LPPA]{Steve N'Guyen}
\author[Sorba,ISIR]{Benoit Girard}

\address[Sorba]{Sorbonne Universit{\'e}s, UPMC Univ Paris 06, UMR 7222, ISIR, F-75005, Paris, France.}
\address[ISIR]{Institut des Systèmes Intelligents et de Robotique, CNRS, UMR 7222, ISIR, F-75005, Paris, France.}
\address[Sorbb]{Sorbonne Universit{\'e}s, Collège de France, UMR 7152, LPPA, F-75005, Paris, France.}
\address[LPPA]{Laboratoire de Physiologie de la Perception et de l’Action, Collège de France, CNRS, UMR 7152, LPPA, F-75231, Paris, France.}

\begin{abstract}
%max 250 words.
The Superior Colliculus, a laminar structure involved in the retinotopic mapping of the visual field, plays a cardinal role in the 
several cortical and subcortical loops of the saccadic system. Although the selection of saccade targets has long been thought to be the sole product of cortical processes, a growing body of evidence hints at the implication of the Superior Colliculus, firstly by the lateral connections between the neurons of its maps, and secondly by its interactions with the midbrain Basal Ganglia, already renowned for their role in decision making. 

We propose a biomimetic population-coded race model of selection based on a dynamic tecto-basal loop that reproduces the observed ability of the Superior Colliculus to stochastically select between similar stimuli, the accuracy of this selection depending on the discriminability of the target and the distractors. Our model also offers an explanation for the phenomenon of Remote Distractor Effect based on the lateral connectivity within the Basal Ganglia circuitry rather than on lateral inhibitions 
within the collicular maps. 

Finally, we propose a role for the intermediate layers of the Superior Colliculus, as stochastic integrators dynamically gated by the selective disinhibition of the Basal Ganglia channels that is consistent with the recorded activity profiles of these neurons.
\end{abstract}

\begin{keyword}
Superior Colliculus \sep saccades \sep selection \sep stochastic race model \sep biomimetic \sep Basal Ganglia
%\MSC[2014] 00-01\sep  99-00
\end{keyword}

\end{frontmatter}

\section{Introduction\label{part_1}}
Saccadic eye movements are probably those for which we make the most decisions. During wakefulness, in ordinary visual conditions, numerous potentially interesting targets constantly compete for further examination by the fovea, and as a few saccades are made each second, decisions concerning the target to be foveated by the next saccade have to be made more than 10⁶ times a day. The cortical circuitry handling these selection processes is now quite well identified, however an increasing number of studies highlight the existence of a possibly autonomous and purely subcortical circuit, also able to select the targets of saccades. We propose here a model of this circuit, and of the neural mechanisms operating the selection and its transformation into an eye motor command.

A common and well-known models for general selection is the race model, in which each competing signal feeds its own evidence counter and the first counter to reach a specific selection threshold gets selected. Another common selection model is the drift-diffusion model, in which a single noisy counter integrates evidence for two competing signals from a common starting point toward one of two selection boundaries (the rate of accumulation, or drift rate, being potentially different for each signal).

Such models have already been proposed to account for cognitive selection processes (\cite{Logan1984}), and particularly for the selection of saccade targets or for saccade coutermanding, with good accuracy (see \cite{Hanes1999,Schall2001,Schall2011} for race models, the latter proposing a gated accumulator able to reproduce Reaction Time recording in easy and difficult discrimination tasks; see \cite{Ratcliff2008} for drift-diffusion models).

Nevertheless, these models are mostly phenomenological, and do not propose a precise or complete role to the various neural structures involved in the specific process of saccade selection. Therefore, their validity can be questioned when confronted to the demands of explaining behavioral resuts with plausible neuronal correlates.

The structures involved in saccade selection processes have been extensively studied, as reviewed in \cite{Moschovakis1996}, and can be broadly summarized as follows: in a first long loop, retinal inputs project to the visual areas of the cortex, and are then processed through the Lateral IntraParietal sulcus (LIP), and the Frontal Eye Fields (FEF). These cortical areas project to several midbrain structures, notably the Basal Ganglia (BG), with which it forms loops also including the Thalamus (Th). The FEF also directly project to the midbrain Superior Colliculus (SC), which itself projects to the brainstem Saccade Burst Generators (SBG) in order to generate a motor command to the Extra-Ocular Muscles (EOM), and to move the eye towards the designed target. 

A second shorter loop sees a direct projection from the retinal inputs to the SC, which then loops with the BG too. The SC ouptut also projects down to the SBG.

From this brief overview, the Superior Colliculus emerges as a central structure, a crossroad between both cortical and subcortical circuits; organized into several dorsoventral layers, its superficial layers receive retinal inputs, while its deeper layers project outputs to orienting motor systems. Between these, intermediary layers combine multisensory integration and premotor activities. These layers are organized in retinotopic maps, from the superficial layers that describe the visual space, to the motor representations of the deep layers that produce ocular orienting responses (see \cite{May2006} for a full review of the SC anatomy). 

Despite the pivotal position of the SC in these loops, target selection in the saccadic system has long been thought to occur at the cortical level - mainly involving the Frontal Eye Fields (\cite{Fischer1987,Schall1998}), while the SC would only serve as a visual mapping structure that relays the selection signal and process the saccade metrics for the brainstem.

\begin{figure*}
%\center\includegraphics[bb=0bp 0bp 1400bp 1150bp,scale=0.45]{fig1/figure_1.pdf}
\center\includegraphics[scale=0.45]{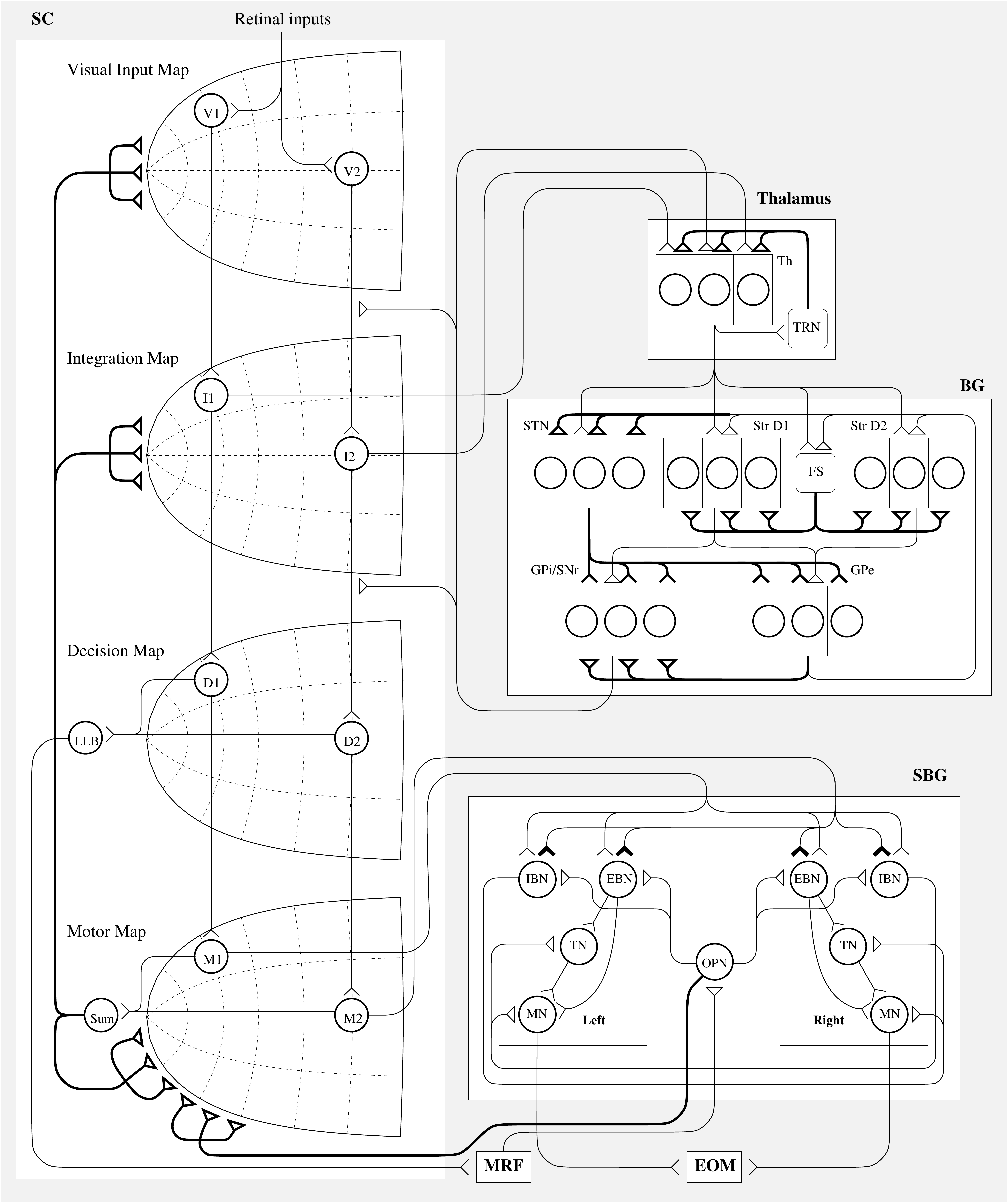}
\caption{\label{fig:model_architecture}{\bf Architecture of the SC-BG model.} Several features have been downscaled for ease of comprehension: only one of the two Colliculi has been represented, and only two neurons in each of the four layer. The architecture of the BG is shown as unidimensionnal when the model represents its channels as a 2D map, and only the projections
from one channel's neurons are represented, since they are identical for the other channels. Only two of the four SBG circuits have been represented, constituting a pair governing eye movements in the horizontal plane. Open triangles represent excitatory synapses, closed triangles represent inhibitory synapses - for the SC-SBG projection only, synapse thickness represents connection strenght. Bold lines represent one-to-all connections, when normal lines represent one-to-one connections. See text for abbreviations.}
\end{figure*}

This view has been challenged, firstly by showing the implication of the SC in saccade target selection (\cite{Mays1980,Ottes1987,Schiller1987}), and secondly by exposing its active role in the selection process (\cite{McPeek2002,McPeek2004,McPeek2008,Carello2004}, among others). The SC laminar organization in several retinotopic neural maps of
the visual field led to the belief that target selection could occur by way of reciprocal lateral short-range excitation and long-range inhibition within each map, as modelled in \cite{Opstal1989} and in other Neural Field models of the SC \cite{Trappenberg2001,Taouali2012}. The winner-takes-all properties of such purely collicular models conflicts with electophysiological and comportemental data regarding the production of averaging saccades (see \cite{Chou1999} for an overview of experimental results concerning averaging saccades): their predictions state that two equal loci of SC activation will merge in a single locus located between the original two, when experimental data shows that the two loci persist (see \cite{Edelman1998}). Moreover, the very existence of longe-range inhibitions has been challenged (\cite{Oezen2004}), as well as the ability of short ranged excitation to explain the effects of proximal distractors on target selection (\cite{Casteau2012}).

Furthermore, such models do not take into account the various connections between the SC and the Basal Ganglia (\cite{Hikosaka1983,McHaffie2005}), a set of midbrain nuclei renowned for its role in action selection (\cite{Mink1996,Ding2013}), especially in the generation of saccades (\cite{Hikosaka2000}). Divided in input structures, the Striatum (Str) and SubThalamic Nucleus (STN), and output structures, the Globus Pallidus external and internal (GPe and GPi), and the Substancia Nigra pars reticulata (SNr), the nuclei of the BG are connected in multiple and identical parallel circuits, included in various parallel cortico-baso-thalamo-cortical loops as well as purely subcortical loops. The subcortical connection between the SC and BG takes the form of loops from the various layers of the superficial and intermediate SC, that project to the BG with a thalamic intermediary. The BG output is then fed back to the same layers. \cite{McHaffie2005,McHaffie2006,Shires2010} have proposed several roles for this BG-SC pathway with regard to vision and action, such as target selection and saccade choice and valuation. 

Thus, we propose a new model, based on the SC model proposed by \cite{Tabareau2007}, and the BG model proposed by \cite{Girard2008}, in which saccade target selection occurs not by way of lateral inhibitions within the SC, but by the race-like competition of target loci and target values mapped in the superficial layers of the SC by direct retinal inputs through the BG, which act as the threshold detector for the evidence accumulated by each target during competition. 

\section{Materials and Methods\label{part_2}}

The model is based on a rate population coding, in which all neurons of a given population are assumed to be sensitive to the same set of inputs and share similar electrophysiological properties. Each of these populations of neurons can therefore be modeled by only one equation, that returns the mean firing rate of the population. The model is organized in three main modules, the SC, BG and SBG, interconnected as shown in fig.~\ref{fig:model_architecture}.

On the mathematical side, in order to unify the theoretical models and equations describing the various neurons of the model, we used the model of locally Projected Dynamical Networks (lPDS) neurons proposed by \cite{Girard2008} for all the neurons of the BG, SC and SBG, unless specifically stated otherwise. The activity of any neuron $x$ under this model, when updated with Euler integration, obeys to the following equation type:
\begin{eqnarray}
a_{x}(t+dt)= & max(0,min(1,a(t)+\nonumber \\
 & \frac{dt}{\tau_{x}}\times(I_{x}(t)-\gamma\times a_{x}(t))))\label{eq:lPDS}
\end{eqnarray}

With $a_{x}(t)$ the activity of the neuron at time $t$, $I_{x}(t)$ the weighted sum of all excitatory and inhibitory inputs at time $t$, $\gamma$ the leak, and $\tau_{x}$ the time constant in ms (both of which being identical for all neurons in the model, unless specified otherwise). 

Note that all the base unit for integration timestep $dt$ is the millisecond.

\subsection{The Superior Colliculus \label{part_2.1}}

The SC module of the model is based on the model proposed by \cite{Tabareau2007} for the description of the deep layers of the SC and the process of Spatio-Temporal Transformation that turns the spatial activity of the whole Motor map of the SC into a temporal signal for the SBG. We also took inspiration from \cite{Tabareau2007} concerning the gluing mechanism that allows the operation of the two colliculi coding each for one visual hemifield into one abstract mapping on the whole plane.

Our SC is divided in two elements, Right and Left, each receiving and processing inputs from the contralateral Retina. Furthermore,
both elements are organized in 4 layers representing retinotopic maps with logarithmic mapping, in order to account for the laminar structure of the SC and the functional properties of the SC layers (see fig.~\ref{fig:model_architecture} for the architecture of one colliculus of the SC module). These layers are composed of $NbCell\times NbCell$ lPDS neurons, and the right
and left colliculi are connected so that the combined activity of their motor layers can be considered as a single "abstract" mapping on the whole plane.

As explained by \cite{Ottes1986}, the logarithmic mapping governing the geometry of the SC transforms the retinotopic Cartesian coordinates ($[az,el]$ for azimuth and elevation) of a target's position in the visual field into coordinates expressed in millimeters ($[X,Y]$) onto the SC surface by the following equation:
\begin{equation}
\frac{X}{B}+i\times\frac{Y}{B}=\ln(\frac{z+A}{A})\label{eq:SC_mapping}
\end{equation}

With $z=az+i\times el$, and A and B experimentally estimated for the monkey to be respectivelly $\pi/60$ and $1.5$.

The most superficial layer is called the Visual layer (or map), and receives direct projections from the retina (cortical inputs are not taken into account in our strictly subcortical model). The retinal input for each target $Targ_{i}$ is represented as a 2D-Gaussian with standard deviation $\sigma=2.5$ neurons (corresponding to $0.35mm$ in the monkey SC) and maximal height equal to the target's value, centered on the neuron at coordinates $[x_{i},y_{i}]$ in our discretized maps, as per equation (2). Furthermore, the Visual map receives an inhibitory projection from the Summation neurons of the Motor layer, which allows the progressive reset of the map according to the execution status of the current saccade (the details of which are described below).

A neuron located at coordinates $x,y$ in the Visual map obeys to equation \ref{eq:lPDS}, with the following inputs:

\begin{equation}
I_{x,y}^{Vis}=I_{x,y}^{Retina}-\omega_{Sum}^{Vis}\times a_{Sum}\label{eq:SC-Vis}
\end{equation}

With $I_{x,y}^{Retina}$ the input from the retina for the same neuron, modulating by a gluing process for vertical or quasi-vertical target positions (details for the calculation of the Retinal input are given in \ref{sec:retinal_input}), and $\omega_{Sum}^{Vis}$ the weight of the inhibitory connection from the the Summation neurons. $\gamma$ for the whole Visual map is set to $1$.

The Visual map projects to the Integration map by one-to-one connections. Each neuron in the Integration map acts as a noisy evidence counter, and integrate the activity of its corresponding Visual map neuron over time, accumulating the value of the target (or targets) exciting said Visual neuron at a variable rate depending on said value. Therefore, the whole Integration map is akin to a multitude of stochastic race models, each acculumating evidence for the selection of one position in the Visual field. 

Furthermore, the Visual map to Integration map connection is subjected to a modulatory inhibition $\Gamma_{BG}^{Int}$ from the BG output, which acts as a gate for selection, allowing for a boost of the integration rate of the selected target and a simultaneous braking of the integration rate of its competitors. Lastly, the Integration map receives the same inhibitory projection from the Summation neurons as the Visual map. Consequently, a neuron located at coordinates $x,y$ in the Integration map obeys to equation \ref{eq:lPDS}, with the following inputs: 
\begin{eqnarray}
I{}_{x,y}^{Int}= & \omega_{Vis}^{Int}\times a_{x,y}^{Vis}\times\Gamma_{BG}^{Int}-\omega_{Sum}^{Int}\times a_{Sum}\nonumber \\
 & +\sqrt{\frac{\tau}{dt}}\times\mathcal{N}(0,\omega_{n}\times\sqrt{a_{x,y}^{Int}}+\varepsilon)\label{eq:SC-Integ}
\end{eqnarray}

\begin{figure}[h]
\includegraphics[scale=0.33]{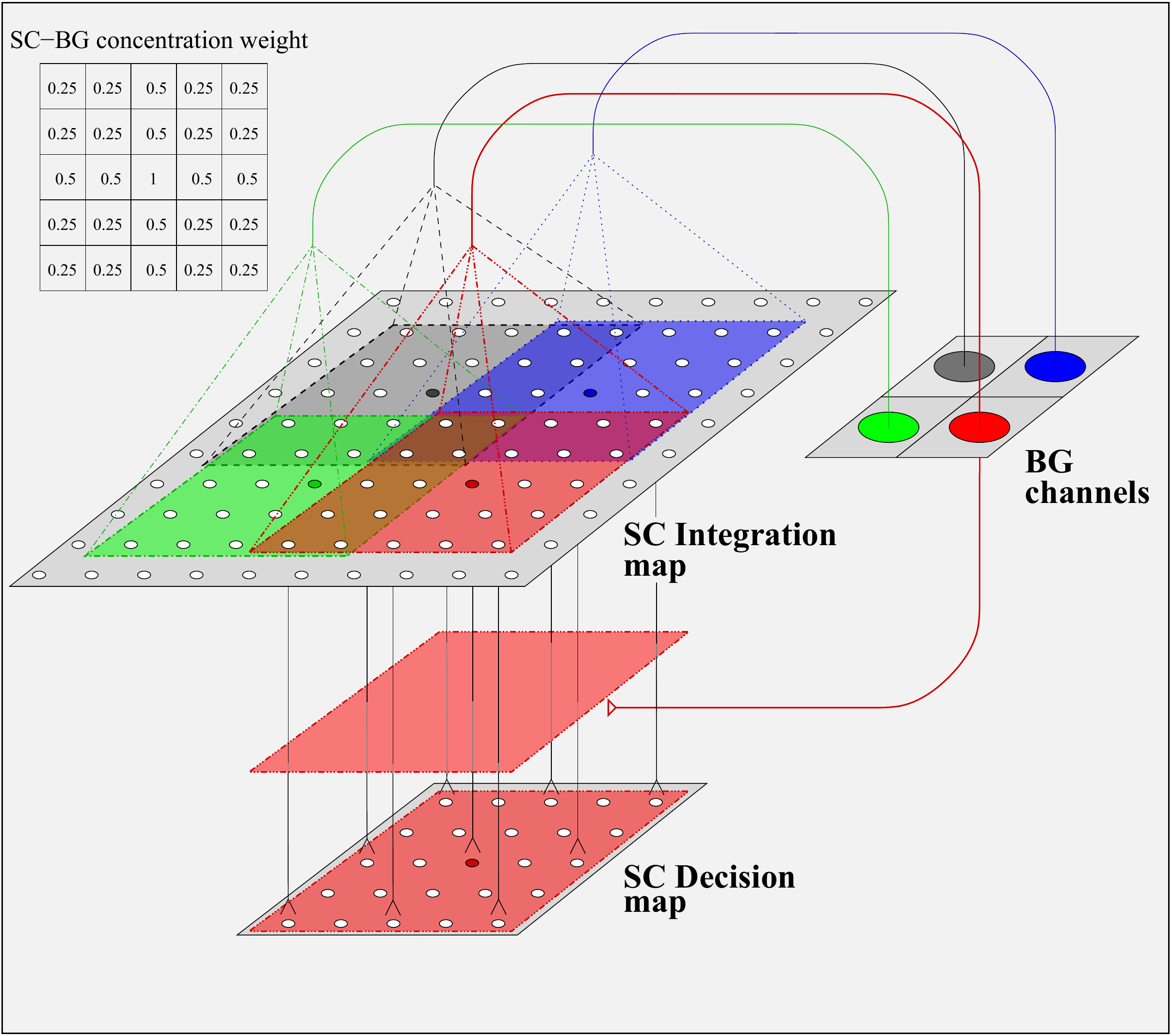}
\caption{\label{fig:SC-BG_connectivity}{\bf Anatomy of the SC-BG connections.} Concentration of the SC integration map output to the BG channels, and deconcentration of the BG output to the connections between the SC maps. The Integration map is divided in subsections of $5\times5$ neurons overlapping each other. Each SC map subsection feeds one specific BG channel with the weight pattern given in the inset: the neuron at the center of the red subsection feeds only one BG channel, while the four neurons of each of its corners all feed four different channels, and the neurons in the middle of the edges of this subsection feed two different channels. BG output to the one-to-one connections between the Integration and Decision maps (and the connection between the Visual and Integration maps, not represented here) obey to the reverse projection pattern, with the same weight pattern so that all BG channels have the same global weigth to the whole SC map.}
\end{figure}

With $\omega_{Vis}^{Int}$ the weight of the input from the Visual map neuron at coordinates $x,y$, $\Gamma_{BG}^{Int}$ the modulatory inhibition exerted by the BG on the connection between the Visual neuron at coordinates $x,y$ and the Integration neuron at the same coordinates (see fig.~\ref{fig:SC-BG_connectivity} for the details of this connectivity), and $\omega_{Sum}^{Int}$ the weight of the inhibitory connection from the the Summation neurons. $\gamma$ for the whole Integration map is set to $0.05$, such low value allowing for the integration of activity over time. 

The gaussian white noise applied to this neuron is proportional to the square root of its previous activity, modulated by a weight $\omega_{n}$. 

This dependency to activity is necessary so that targets close to the vertical have the same probability of being selected than targets elsewhere (see the gluing description in \ref{sec:retinal_input}).

The modulatory inhibition $\Gamma_{BG}^{Int}$ is calculated for each SC Integration map neuron as per equation \ref{eq:SC-BG_inhib-integ}:
\begin{equation}
\Gamma_{BG}^{Int}=1-\frac{{\displaystyle \sum_{x,y\in N}}a_{x,y}^{BG}}{T_{BG}^{Int}}\label{eq:SC-BG_inhib-integ}
\end{equation}

With $T_{BG}^{Int}$ the specific threshold for the basal output of the BG to the Integration map, and ${\displaystyle \sum_{x,y\in N}}a_{x,y}^{BG}$ the summed outputs of all channels of coordinates $x,y$ in the submap $N$ of the BG, feeding one given SC Integration map neuron (see fig.~\ref{fig:SC-BG_connectivity} for the details of this connectivity).

The threshold $T_{BG}^{Int}$ is set to 0.359, higher than the BG resting output (mesured at 0.349), with the effect of allowing some communication between the Visual and Integration map even when the BG inhibitory output is at its basal firing rate. When this output changes from its rest level, $\Gamma_{BG}^{Int}$ will either enhance or decrease the weight of each individual connection (and therefore the rate of integration of each Integration neuron) according to the variation of the output of the channels fed by this neuron: BG channels coding for a target that gets little evidence counted in the integration map will lose selection to channels coding for a target that gets more evidence counted, and therefore the "losing" BG channels output will have a stronger inhibitory influence on their Integration neurons than at rest level, when "winning" channels will have a weaker inhibitory influence.

Thus, the rate of integration of each Integration neuron will be enhanced if this neuron codes for a target in the process of being selected, but decreased if the neuron codes for a target in the process of not being selected.

The Integration map output is projected in two different directions: the first one is a one-to-one projection to the Decision map located deeper in the SC, and the second one is aimed at the Thalamus, in order to feed the BG module of the model. This Integration map to BG projection is subjected to a rescaling process in order to account for the differences in dimensions between the SC retinotopic maps and the BG channel maps. The visual field represented by the retinotopic properties of the Integration map is divided in subsections, and a channel of the BG will be dedicated to receiving the concentrated outputs of all neurons within each subsection (these outputs are weighted so that all neurons are used to the same proportions by the whole BG - see fig.~\ref{fig:SC-BG_connectivity} for the details of this connectivity). 

The resulting competition between the BG channels will disinhibit only the specific subsection of the retinotopic map corresponding
to the winning channel, hence the sum term in equ. \ref{eq:SC-BG_inhib-integ}.

The Decision map receives one-to-one inputs from the Integration map, and this connection is also subjected to a modulatory inhibition $\Gamma_{BG}^{Dec}$ from the BG (see fig.~\ref{fig:SC-BG_connectivity} for the details of this connectivity), as per equation \ref{eq:SC-BG_inhib-Decis}:
\begin{equation}
\Gamma_{BG}^{Dec}=1-\frac{{\displaystyle \sum_{x,y\in N}}a_{x,y}^{BG}}{T_{BG}^{Dec}}\label{eq:SC-BG_inhib-Decis}
\end{equation}

With $T_{BG}^{Dec}$ the specific threshold for the basal output of the BG to the Decision map, and ${\displaystyle \sum_{x,y\in N}}a_{x,y}^{BG}$ the summed outputs of all channels of coordinates $x,y$ in the submap $N$ of the BG, feeding one given SC Decision map neuron.

his modulatory inhibition is very similar to the one exerted on the connection between the SC Visual and Integration map with a single, yet major, difference concerning the tuning of the threshold $T_{BG}^{Dec}$: it is set to 0.349 (when $T_{BG}^{Int}$ is set to 0.359), at the same level as that of the BG resting output (0.349), with the effect of completly shutting down the communication between the Integration and Decision map when the BG activity is at its rest level.

When the Integration map has fed enough evidence to the BG for a channel to be selected, the inhibition on the Integration map to Decision map connection will be selectively lifted for the subsection of the retinotopic map corresponding to the winning BG channel, and only the signal generated by the winning target will be transmitted from the Integration map to the Decision map. Thus, the losing targets are erased from this map (while still present in the upper maps), and only the signal related to the winning target is transmitted to the Motor map.

Therefore, a neuron located at coordinates $x,y$ in the Decision map obeys to equation \ref{eq:lPDS}, with the following inputs: 
\begin{equation}
I_{x,y}^{Dec}=(\omega_{Int}^{Dec}\times a_{x,y}^{Int}\times\Gamma_{BG}^{Dec})\label{eq:SC-Decis}
\end{equation}

With $\omega_{Int}^{Dec}$ the weight of the input from the Integration map neuron at coordinates $x,y$, $\Gamma_{BG}^{Dec}$ the modulatory inhibition exerted by the BG on the connection between the Integration neuron at coordinates $x,y$ and the Decision neuron at the same coordinates. $\gamma$ for this map is set to $1$. 

Finally, the neurons of the Motor map receive one-to-one connections from the Decision map, and two inhibitions inherited from the SC model of \cite{Tabareau2007}. The first is exerted by the OPNs of the SBG and ensures that no motor activity is generated while target selection is ongoing. The second inhibition comes from the summation-integration neurons in order to perform the process of Spatio-Temporal Transformation as proposed by \cite{Groh2001}, and ensures that no matter the particularities
of the current saccade profile (from a quick and strong burst of motor activity to a longuer but weaker burst of activity), the integration of the whole Motor map activity remains constant over the duration of the saccade.

A neuron located at coordinates $x,y$ in the Motor map obeys to equation \ref{eq:lPDS}, with the following inputs: 
\begin{equation}
I_{x,y}^{Mot}=\omega_{Dec}^{Mot}\times a_{x,y}^{Dec}\times(1-\omega_{Sum}^{Mot}\times a_{Sum})-\omega_{OPN}^{Mot}\times a_{OPN}\label{eq:SC-Motor}
\end{equation}

With $\omega_{Dec}^{Mot}$ the weight of the input from the Decision map neuron at coordinates $x,y$, $\omega_{Sum}^{Int}$ the weight of the inhibitory connection from the the Summation neurons, and $\omega_{OPN}^{Mot}$ the weight of the inhibitory connection from the OPNs of the SBG. $\gamma$ for the whole Motor map is set to $1$. 

The Motor map's whole activity is then transmitted as a weighted sum to the EBNs and IBNs of the SBG in order to transform the SC spatial information about the saccade target location into a Cartesian temporal motor command that will be transmitted to the EOM in order to move the eye towards the target. The projection weights from the SC to the SBG is tuned to the position in the visual field coded by each Motor map neuron, and also to the horizontal or vertical direction of movement coded by each SBG circuit.

As mentionned earlier, the proper preparation of the motor command as supposed by the process of STT dictates that the activity of the Motor map is bound by two set of neurons, ensuring that no motor command is prepared as long as selection is not achieved, and that the total motor activity over time is constant for all saccade metrics, as hypothetized by \cite{Groh2001}. The LLBs/cMRF and Summation neurons are the operators for these two respective functions.

The LLB of the SC integrates the activity of some of its maps, and excites the cMRF that in turn inhibits the OPNs of the SBG in order to lift the OPNs' basal suppression of the Motor map activity (\cite{Wang2013}). In our model, this LLB/cMRF\ensuremath{\slash}OPN connection is simplified in a direct LLB/OPN inhibitory projection. The LLB are fed by the most superficial map in the SC on which only the activity related to the target(s) having won selection is displayed, that is the Decision map.

The LLBs obey to equation \ref{eq:lPDS}, with the following inputs: 
\begin{equation}
I_{LLB}=\omega_{Dec}^{LLB}\times\sum_{x,y\in SC}a_{x,y}^{Dec}-E_{LLB}\label{eq:SC-LLB}
\end{equation}

With $\omega_{Dec}^{LLB}$ the weight of the Decision map input to the LLB, and $E_{LLB}$ the threshold triggering LLB activity. $\gamma$ for the LLBs is set to $1$. 

The Summation neurons implement the mechanism of summation with saturation as proposed by \cite{Groh2001}: they receive the summed activity of the whole Motor in order to gradually inhibit the SC output, that is the Motor map activity. This mechanism of SC output regulation has been extended in our model to the regulation of the activity of the Visual and Integration maps too.

The Summation neurons are perfect resetable integrators, thus $\gamma=0$ for them. They obey to the following equation: 
\begin{equation}
I_{Sum}=\omega_{Mot}^{Sum}\times\sum_{x,y\in SC}a_{x,y}^{Mot}\label{eq:SC-Sum}
\end{equation}

With $I_{Sum}$ the global input to the Summation neurons, and ${\displaystyle \sum_{x,y\in SC}}a_{x,y}^{Mot}$ the weigted sum of the activity of all Motor map neurons. 

\subsection{The Basal Ganglia\label{part_2.2}}
he BG module of our model is mostly copied from \cite{Girard2008}, with three key features changed:
\begin{enumerate}
\item All the cortical elements of the original model have been removed. Therefore, the Thalamus of our model is only regulated by a global inhibition of the Thalamic Reticular Nucleus and a channel-specific selective inhibition from the BG. 
\item The Thalamus and TRN elements of our BG module are not only collecting the output of the BG, as in \cite{Girard2008}, but are also feeding the input nuclei of the BGs, as indicated by the anatomy of the subcortical BG loops (\cite{McHaffie2005}). Thus, the Thalamus receives the afferences from the SC, and feeds them to the input nuclei of the BG (that is, the Striatum D1 and D2, the FS neurons and the STN).
\item The dimensions and parameters of the BG module have been adaptated to the size of its SC inputs. The channel architecture of each BG nucleus is therefore bidimensionnal, sized to one third of the dimensions of the SC inputs, and its set of parameters is given in Table~\ref{tab_param_CBG}, in appendix \ref{sec:tables_parameters}.
\end{enumerate}

These changes taken into account, our BG module operates in a similar way with the BG model proposed by \cite{Girard2008}: each channel receives inputs (representing the summed activity of the whole subsection of the SC map each channel is connected to), and competes with the other channels in an off-center/on-surround inhibition system that promotes the disinhibition of the winning channel, and the overinhibition of the losing channels. The feedback loop to the SC will reflect this change in the inhibitory output of the BGs, and selectively allow the transmission of signal from one map of the SC to the other only for the subsections linked to the winning channel of the BGs (see fig. ~\ref{fig:model_architecture} for the anatomy of the BG module, and fig.~\ref{fig:SC-BG_connectivity} for the detailled description of the connectivity between the BG channels and SC maps).

The input nuclus of our BG module is the Th, which is divided in channels. It receives projection from the Integration map of the SC, as well as a regulatory diffuse inhibition from the Thalamic Reticular Nucleus (TRN), and a specific channel-to-channel feedback inhibition from the output nucleus of the BG, the GPi.

Each channel $i$ obeys to equation \ref{eq:lPDS}, with the following inputs:
\begin{eqnarray}
I_{i}^{Th} & = & \omega_{SC}^{Th}\times\sum_{x,y\in SC}a_{x,y}^{SC}-\omega_{TRN}^{Th}\times a_{TRN}\nonumber \\
 &  & -\omega_{BG}^{Th}\times a_{i}^{GPi}+E_{Th}\label{eq:BG-Th}
\end{eqnarray}

With $\omega_{SC}^{Th}$ the weight of the input from the subsection of the SC Integration map of coordinates $x,y$ corresponding to channel $i$, $\omega_{TRN}^{Th}$ the weight of the TRN input to the Th, $\omega_{BG}^{Th}$ the weight of the GPi input from channel $i$, and $E_{Th}$ the basal activity of the Th. $\gamma$ for each channel of the Th is set to $1$. 

The TRN is not organised into channels, but aggregates activity from all Th channels in order to exert to a global feedback inhibition to the Th. It obeys to equation \ref{eq:lPDS}, with the following inputs:
\begin{equation}
I_{TRN}=\omega_{Th}^{TRN}\times\sum_{i\in BG}a_{i}^{Th}\label{eq:BG-TRN}
\end{equation}

With $\omega_{Th}^{TRN}$ the weight of the Th input to the TRN. $\gamma$ for the TRN is set to $1$ (the TRN time constant is $\tau_{small}$ rather than the standard $\tau$). 

The BGs themselves are composed of several nuclei: the Striatum, with its D1 and D2 Medium Spiny Neuron populations plus the Fast Spiking (FS) interneurons, and the STN - the two of them constituting the input structures of the BGs, the intermediary GPe, and the GPi/SNr, the output structure of the BGs.

The FS neurons of the Striatum, like the TRN, are represented by one single population of neurons rather than being divided in channels. They exert a feedforward inhibition on the MSN. They are fed by the summed output of the Th, and regulated by the summed inhibitory output of the GPe. They obey to equation \ref{eq:lPDS}, with the following inputs:
\begin{equation}
I_{FS}=\omega_{Th}^{FS}\times\sum_{i\in BG}a_{i}^{Th}-\omega_{GPe}^{FS}\times\sum_{i\in BG}a_{i}^{GPe}\label{eq:BG-FS}
\end{equation}

With $\omega_{Th}^{FS}$ the weight of the Th input to the FS, and $\omega_{GPe}^{FS}$ the weight of the GPe input to the FS. $\gamma$ for the FS is set to $1$, and its time constant is $\tau_{small}$ rather than the standard $\tau$. 

The other striatal neuron populations represented in the model are the D1 and D2 MSNs. They are divided in channels, and obey both to the same global equation \ref{eq:lPDS}, with channel-to-channel Th excitation, GPe inhibitory feedback and global FS inhibitory feedback; D1 and D2 differ only in the effect of the dopamine level: excitatory for D1 and inhibitory for D2.
\begin{eqnarray}
I_{i}^{D1}=(1+\lambda)\times & (\omega_{Th}^{D1\times}a_{i}^{Th}-\omega_{GPe}^{D1}\times a_{i}^{GPe})\nonumber \\
 & -\omega_{FS}^{D1}\times a_{FS}+E_{D1}\label{eq:BG-D1}
\end{eqnarray}

\begin{eqnarray}
I_{i}^{D2}=(1-\lambda)\times & (\omega_{Th}^{D2}\times a_{i}^{Th}-\omega_{GPe}^{D2}\times a_{i}^{GPe})\nonumber \\
 & -\omega_{FS}^{D2}\times a_{FS}+E_{D2}\label{eq:BG-D2}
\end{eqnarray}

With $\lambda$ the dopamine level modulating the inputs to the dendritic tree of either MSN population (kept constant in this study), $\omega_{Th}^{D1/D2}$ and $\omega_{GPe}^{D1/D2}$ the weights of the inputs from the Th and the GPe respectively, $\omega_{FS}^{D1/D2}$ the weight of the FS input to the MSNs. Finally, the negative constant inputs $E_{D1/D2}$ keep the neurons silent when the thalamic inputs are not strong enough, and represent the up-state/down-state property of these neurons. $\gamma$
for both D1 and D2 MSN populations is set to $1$.

The STN is the second input structure of the BGs, and is also divided in channels. It receives channel-to-channel inputs from the Th, as well as a global GPe inhibitory feedback. Each of its channels $i$ obeys to equ. \ref{eq:lPDS}, with the following inputs:
\begin{equation}
I_{i}^{STN}=\omega_{Th}^{STN}\times a_{i}^{Th}-\omega_{GPe}^{STN}\times\sum_{i\in BG}a_{i}^{GPe}+E_{STN}\label{eq:BG-STN}
\end{equation}

With $\omega_{Th}^{STN}$ the weight of the Th input to the STN, $\omega_{GPe}^{STN}$ the weight of the summed GPe input to the STN, and $E_{STN}$ the basal activity of the STN. $\gamma$ for the STN is set to $1$, and its time constant is $\tau_{small}$ rather than the standard $\tau$. 

The GPe is an intermediary inhibitory nucleus of the BGs, that receives channel-to-channel inhibitory inputs from the D1 and D2 populations of the Striatum, and a diffuse excitation from the STN. Each channel $i$ of the GPe obeys to equation \ref{eq:lPDS}, with the following inputs:
\begin{eqnarray}
I_{i}^{GPe} & = & \omega_{STN}^{GPe}\times\sum_{i\in BG}a_{i}^{STN}-\omega_{D1}^{GPe}\times a_{i}^{D1}\nonumber \\
 &  & -\omega_{D2}^{GPe}\times a_{i}^{D2}+E_{GPe}\label{eq:BG-GPe}
\end{eqnarray}

With $\omega_{STN}^{GPe}$ the weight of the summed STN input to the GPe, $\omega_{D1/D2}^{GPe}$ the weight of the input from either D1 or D2 MSNs to the GPe, and $E_{GPe}$ the basal activity of the GPe. $\gamma$ for each channel of the GPe is set to $1$.

he GPi/SNr is the output nucleus of the BGs. It receives a diffuse excitation from the STN, a diffuse inhibition form the GPe, and channel-to-channel inhibitory inputs from the D1 and D2 populations of the Striatum. Each channel $i$ of the GPi obeys to equation \ref{eq:lPDS}, with the following inputs:
\begin{eqnarray}
I_{i}^{GPi} & = & \omega_{STN}^{GPi}\times\sum_{i\in BG}a_{i}^{STN}-\omega_{GPe}^{GPi}\times\sum_{i\in BG}a_{i}^{GPe}\nonumber \\
 &  & -\omega_{D1}^{GPi}\times a_{i}^{D1}-\omega_{D2}^{GPi}\times a_{i}^{D2}+E_{GPi}\label{eq:BG-GPi}
\end{eqnarray}

With $\omega_{STN}^{GPi}$ the weight of the summed STN input to the GPi, $\omega_{GPe}^{GPi}$ the weight of the summed GPe input to the GPi, $\omega_{D1/D2}^{GPi}$ the weight of the input from either D1 or D2 MSNs to the GPi, and $E_{GPi}$ the basal activity of the GPi. $\gamma$ for each channel of the GPi is set to $1$.

\subsection{The Saccade Burst Generators\label{part_2.3}}

The SBG module of the model is reproduced from \cite{Tabareau2007}, with some minor parameters adjustments in order to account for the changes in the activity profiles of the Motor layer of our model.

The SBG is composed of four identical circuits, each responsible for the rotation of the eye in either the Rightward, Leftward, Upward or Downward direction ($dir\in\left[R,L,U,D\right]$). These circuits are coordinated in pairs by crossed projections from some of their input neurons to the output neurons of the coordinated circuit, operating in opposed directions $dir$ and $dir_{opp}$ (see fig.~\ref{fig:model_architecture} for the details of the SBG anatomy and the connectivity between the circuits in one pair).

Each circuit is constituted of a population of Excitatory Burst Neurons, a population of Inhibitory Burst Neurons, a population of Tonic Neurons and a population of MotoNeurons. There is only one set of Omni-Pause Neurons for the whole SBG, that projects to each of the four circuits and gates their activity. 

In each circuit, movements are encoded by bursts of activity representing the vectorial components of the desired rotation, produced as follows:

The OPNs are tonically active, and exert a basal inhibition that completely shuts down the activity of the input neurons of the SBG, the EBNs and IBNs as well as the whole Motor map of the SC. This inhibition prevents the generation or transmission of unwanted signal through the SBG, and the generation of parasitic eye movements. The OPNs are themselves inhibited by the LLBs of the SC when the activity in the Decision layer of the SC is high enough to cross the LLBs activation threshold: 
\begin{equation}
I_{OPN}=-\omega_{LLB}^{OPN}\times a_{LLB}+E_{OPN}\label{eq:SBG-OPN}
\end{equation}

With $\omega_{LLB}^{OPN}$ the weight of the LLB input to the OPNs, and $E_{OPN}$ the basal activity of the OPNs. $\gamma$ for the  OPNs is set to $1$.

The EBNs and IBNs are basally shut down by the OPNs. They also receive the output of the whole Motor layer of the SC, weighted accordingly to the horizontal or vertical direction of movement relevant to each circuit.

Therefore, the inputs for the EBNs of the SBG circuit coding for direction $dir$ answer to the following equation:
\begin{equation}
I_{EBN}^{dir}=\sum_{x,y\in SC}(\omega_{SC}^{SBG-dir}\times a_{x,y}^{Mot})-\omega_{OPN}^{BN}\times a_{OPN}\label{eq:SBG-EBN}
\end{equation}

With $\omega_{SC}^{SBG-dir}$ the weight pattern of the SC-SBG projection for direction $dir$, as defined by equ.3 of \cite {Tabareau2007} and represented in supplementary fig.~\ref{fig:annexe2_SC-SBG_weights}, and $\omega_{OPN}^{BN}$ the weight of the OPN inhibition on the EBNs. $\gamma$ for the EBNs is set to $1$.

The IBNs have exactly the same inputs, characteristics and mathematical properties as the EBNs, and differ only in the effects of their outputs on their targets : the IBNs of the SBG circuit coding for direction $dir$ will inhibit the TN and MN of the SBG circuit coding for the opposed direction $dir_{opp}$: the IBNs of the Upward SBG circuit will feed the TNs and MNs of the Downward circuit, while the IBNs of the Downward SBG circuit will feed the TNs and MNs of the Upward circuit. The same scheme occurs between the Rightward and Leftward circuits.

The TNs of the SBG circuit coding for direction $dir$ have a non-zero resting activity, and integrate the difference between the output of EBN from the same SBG circuit and the output of the IBN of the SBG circuit coding for the opposed direction $dir_{opp}$:
\begin{equation}
I_{TN}^{dir}=\omega_{BN}^{TN}\times(a_{EBN}^{dir}-a_{IBN}^{dir_{opp}})\label{eq:SBG-TN}
\end{equation}

With $\omega_{BN}^{TN}$ the weight of the connection from the EBNs and IBNs to the TNs , $a_{EBN}^{dir}$ the activity of the EBNs of the SBG circuit coding for direction $dir$, and $a_{IBN}^{dir_{opp}}$ the activity of the IBNs of the SBG circuit coding for the opposed direction $dir_{opp}$. The TNs are perfect integrators, thus $\gamma=0$ for them.

The MNs of the SBG circuit coding for direction $dir$ receive inputs from the EBNs and TNs of their own SBG circuit, and from the IBNs of the SBG circuit coding for the opposed direction $dir_{opp}$. They have a non-zero basal activity due to the non-zero  resting TN component of their input, and therefore always produce a motor command towards the EOM, that allows for the stability of gaze between movements. 
\begin{equation}
I_{MN}^{dir}=\omega_{BN}^{MN}\times(a{}_{EBN}^{dir}-a_{IBN}^{dir_{opp}})+\omega_{TN}^{MN}\times a_{TN}^{dir})\label{eq:SBG-MN}
\end{equation}

With $\omega_{BN}^{MN}$ the weight of the connection from the EBNs and IBNs to the MNs , $a_{EBN}^{dir}$ the activity of the EBNs of the SBG circuit coding for direction $dir$, and $a_{IBN}^{dir_{opp}}$ the activity of the IBNs of the SBG circuit coding for the opposed direction $dir_{opp}$, and $\omega_{TN}^{MN}$ the weight of the TNs projection to the MNs. $\gamma$ for the MNs is set to $1$.

Finally, when activity in the SC Decision layer decreases under the LLBs activation threshold, the OPNs suppression will progressively be lifted and the OPN will start inhibiting the EBNs and IBNs again, thus ensuring that the motor command to the EOM is limited in time. 

The biomecanics accounting for the eye movement are given by a standard second-order differential equation that link the movement of the eye in the horizontal or vertical directions to the difference between the firing rates of the MNs from the SBG circuits coding for this direction and the opposed one: 
\begin{equation}
\ddot{\theta}+\omega_{\theta_{vit}}^{\theta_{acc}}\times\dot{\theta}+\omega_{\theta_{pos}}^{\theta_{acc}}\times\theta=\omega_{MN}^{\theta_{acc}}\times(a_{MN}^{dir}-a_{MN}^{dir_{opp}})\label{eq:SBG-eyeplant}
\end{equation}

With $\theta$ the position of the eye, $\dot{\theta}$ its movement speed, $\ddot{\theta}$ its movement acceleration, $a_{MN}^{dir}$
and $a_{MN}^{dir_{opp}}$ the activity of the MNs of the SBG circuit coding for direction $dir$ and its opposite $dir_{opp}$, and $\omega_{MN}^{\theta_{acc}}$ the weight of the connection between the MNs of the SBG and the eye plant.

\subsection{Model parameterization\label{part_2.4}}

All modules (SC, BG and SBG) of the model have been hand-tuned individually in order to assess their proper functioning before being linked together. Then, further tuning was made so that the whole model would perform correctly.

The SBG module kept the original parameters from \cite{Tabareau2007}, except for the weight $\omega_{MN}^{\theta_{acc}}$, which sets the gain of the saccades. This parameter was hand-tuned in order to produce saccades with correct amplitudes.

Since the number of channels in the BG module is much higher here than in \cite{Girard2008} (121 compared to 6), the strength of all diffuse connections within the module had to be tuned down in order to prevent diffuse connections from always shutting down any one-to-one connection, and to allow the selective disinhibition of one channel. To reach this goal, the isolated BG module was fed with "targets" modelled by 2D Gaussian inputs similar to those used in the tasks, with varied values. The parameters were adjusted until the selection of a single target with an value between 0.6 and 1 was restored. Finer adjustments were then made so that one or two distractors of inferior values would not disturb the selection process, and that the simultaneous selection of multiple targets occurred only when they have very close values. 

The SC module has no significant changes in parameters for all features directly reproduced from \cite{Tabareau2007}, such as the motor layer and integration-saturation mechanisms. The parameters for the added or heavily modified layers, such as the Visual, Integration and Decision map are mostly similar with those of the Motor layer, being based upon the same equation types. This module contains three critical parameters that had to be tuned with care, in order to have an optimal compromize between the duration of the selection process (whether the competing targets' values are similar or not), the production of accurate saccades, the minimization of the average-to-normal saccades ratio, and the production of realistic activity patterns for the various
neurons and maps they affect: 
\begin{itemize}
\item the noise weight $\omega_{n}$ (cf. equ \ref{eq:SC-Integ}). This parameter has to bet set low enought so that the noise level remains low compared to the target's value, but high enough so that the duration of stochastic discrimination between two identical targets by the BG remains compatible with the normal latency for saccades (under 100ms for fast saccades with "easy" selection choices, and up to 200ms for saccades with harder selection choices). Thus, it was defined using a grid search on the $[0,1]$ interval, by steps of 0.05.
\item the threshold $T_{BG}^{Int}$ (described in equ \ref{eq:SC-BG_inhib-integ}) for the BG feedback to the Visual-to-Integration maps connection. This parameter has to be higher that the BG resting output to allow the Visual map to Integration map connection, but not so high that it would take too long for the integrators to feed enough evidence to the BG to reach this threshold within the realistic timeframes mentionned earlier. This was achieved by grid-searching for the value yielding the best results within the $[0.350,0.365]$ interval, by steps of 0.001. 
\item the threshold $T_{BG}^{Dec}$ (described in equ \ref{eq:SC-BG_inhib-Decis}) for the BG feedback to the Integration-to-Decision maps connection, that allows the BG to basally inhibit all connections between the Integration map and the Decision map but, once selection is achieved, this threshold allows one-to-one connections to the Decision map only from the neurons of the Integration map coding for the winning target's location; therefore, this parameter needs to be lower than the BG resting output, but still high enough so as to minimize the occurence of cases where more than one target location is gated from the Integration map to the Decision map. This was achieved by grid-searching for the value yielding the best results within the $[0.340,0.350]$ interval, by steps of 0.001. 
\end{itemize}

\subsection{Simulated tasks\label{part_2.5}}

\begin{table*}[t]
\center
\begin{tabular}{|c|c|c|c|c|c|c|}
\hline 
condition & position 1 & position 2 & position 3 & position 4 & position 5 & position 6\tabularnewline
\hline 
1 point & {[}$20\,^{\circ}$,$20\,^{\circ}${]} & - & - & - & - & -\tabularnewline
\hline 
2 points & {[}$20\,^{\circ}$,$20\,^{\circ}${]} & {[}$-20\,^{\circ}$,$-20\,^{\circ}${]} & - & - & - & -\tabularnewline
\hline 
4 points & {[}$20\,^{\circ}$,$20\,^{\circ}${]} & {[}$20\,^{\circ}$,$-20\,^{\circ}${]} & {[}$-20\,^{\circ}$,$-20\,^{\circ}${]} & {[}$-20\,^{\circ}$,$20\,^{\circ}${]} & - & -\tabularnewline
\hline 
6 points & {[}$20\,^{\circ}$,$0\,^{\circ}${]} & {[}$10\,^{\circ}$,$-10{*}\sqrt{3}\,^{\circ}${]} & {[}$-10\,^{\circ}$,$-10{*}\sqrt{3}\,^{\circ}${]} & {[}$-20\,^{\circ}$,$0\,^{\circ}${]} & {[}$-10\,^{\circ}$,$10{*}\sqrt{3}\,^{\circ}${]} & {[}$10\,^{\circ}$,$10{*}\sqrt{3}\,^{\circ}${]}\tabularnewline
\hline 
\end{tabular}
\caption{\label{tab_positions-McPeek} {\bf Point positions in the visual field the various conditions of task 3.} The total number of points in the visual field for each test condition is given in the first column. Coordinates are given in $[azimuth,elevation]$ in the following columns. Eccentricity is constant across all setups for all elements, and all elements are equidistant from each other in each setup. The target is always located at position 1 in the target-and-distractors setup.}
\end{table*}

Targets are characterized by three parameters: their coordinates, given by their azimuth and elevation in the Cartesian coordinates
of the visual field projecting on the superficial visual layer of the SC; and their value, indicating the height of the 2D Gaussian
representing the target in the model. 

Azimuth and elevation both vary in the $[-30\,^{\circ},30\,^{\circ}]$ range in Cartesian coordinates, and target value is scaled on $[0,1]$.

Unless specified otherwise, any simulation occurs as follows: 
\begin{itemize}
\item the model is initiated for 30 ms so that all of its components reach stable basal activity levels.
\item targets are presented accordingly to each task protocol.
\item the simulation runs its course until either 250 ms after a saccade has been produced, or 750 ms after targets presentation if no saccade has been produced.
\end{itemize}

\subsubsection{Task 1: model characterization \label{part_2.5.1}}

In this task, the model produces saccades towards a single target, the position of which varies over the full breadth of the simulated visual field. This will allow for the assessment of the precision of saccade generation and the production of control datasets for all targets positions when no selection occurs.

Another goal of this task is to ensure that the activity profiles of the various neurons of the model respect the electrophysiological properties of their in-vivo counterparts, as recorded in the litterature.

Target value is always maximal, and a hundred tests are launched for each set of target coordinates, which both vary by increment of $1\,^{\circ}$ in the $[-30\,^{\circ},30\,^{\circ}]$ range.

\subsubsection{Task 2: selection between targets and distractors, or between distractors only \label{part_2.5.2}}

This task aims at reproducing the task of discrimination between a target and a various number of distractors used by \cite{McPeek2002,McPeek2004,McPeek2008}. In this task, a monkey is trained to discriminate between one target and several distractors, and to produce a saccade towards the target. This is tested with or without the injection of Lidocaïne or Muscimol in the SC at the location coding for the position of the target, in order to study the role of the SC in saccade target selection.

We reproduce the effect of training the monkey for the task by setting the target's value at twice that of the distractors (with target value varying between 0 and 1 by increments of 0.05, and therefore discriminators value varying between 0 and 0.5 by increments of 0.025), and we test the ability of the model to perform accurate selection between one target and a variable number of distractors. We test four visual field setups with a varying number of points in the Retinal input, as described in Table~\ref{tab_positions-McPeek}, four hundred times each, with the target always located at the coordinates given for position 1. This control setup is called the "target-and-distractors" setup. 

The test element of the experimental task uses the same setup as the control task, but with the addition an injection of Lidocaïne/Muscimol at the SC location coding for the targets's coordinates, dosed in such way as to decrease local neuronal activity without completely shutting it down. Thus, it is hoped that the increased value the monkey learned to place on the target will be offset by the reduced neuronal activity at the target locus in the SC, and discrimination between target and distractors should be more  difficult, the target being more or less considered as another distractor.

We reproduce the effects of the drug-injection by lowering the maximal value reachable by neurons at the injection site to the level of the value of distractors, all points in each condition having therefore the exact same value varying between 0 and 1 by increments of 0.05, in the same four visual fields conditions as the control condition - each being tested four hundred times as well. This test setup is called the "distractors-only" setup.

The difficulty for the model in this setup lies in its ability in selecting only one distractor among many identical ones, in a finite time, and without resulting in too many average saccades which are the symptoms of the simultaneous selection of multiple distractors.

\subsubsection{Task 3: effects of the separation of targets on the production of accurate saccades \label{part_2.5.3}}

This task is derived from the 2 elements condition of the distractors-only setup of task 2: it aims at characterizing the limits of the model in discriminating between two competing targets of similar value, considering the separation between the targets. We test the model's ability to produce a good balance of correct and average saccades when presented with one target T1 at a fixed position and another target T2 of similar value but variable position.

\begin{figure}[!h]
\includegraphics[scale=0.75]{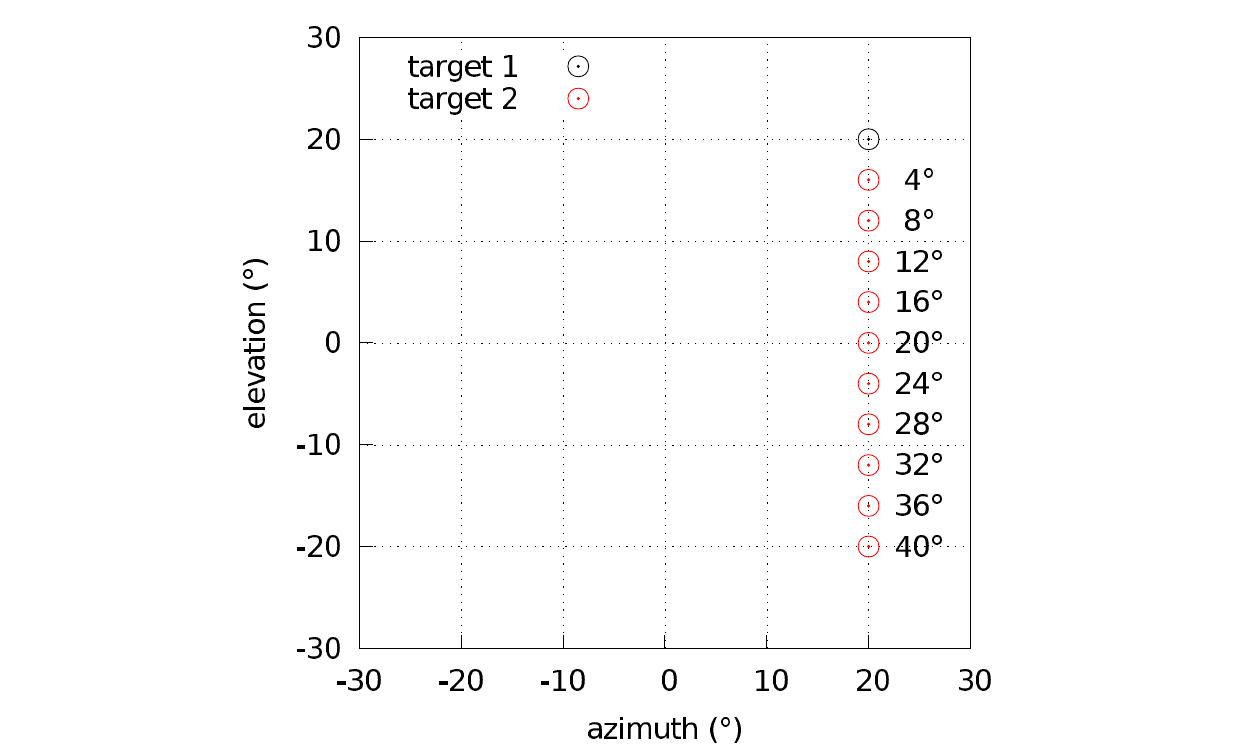}
\caption{\label{fig:metric_2C_positions}{\bf Sample of target positions for task 3.} Target 1 (black circle) is always located at coordinates $[20\,^{\circ},20\,^{\circ}]$, and target 2 (red circle) is located at a variable distance from target
1 on the same vertical plane. Also noted is the distance between the two targets.}
\end{figure}

The two targets are displayed simultaneously after model initialization, with the following parameters:
\begin{itemize}
\item T1 has a value of 1 and is located at coordinates $[20\,^{\circ},20\,^{\circ}]$.
\item T2 also has a value of 1; its azimuth is set to $20\,^{\circ}$ and its elevation varies in the $[-20\,^{\circ},20\,^{\circ}]$ range by increments of $1\,^{\circ}$ (see fig.~\ref{fig:metric_2C_positions}).
\end{itemize}
Each condition is tested two hundred times.

\section{Results\label{part_3}}

Before anything else, two criteria concerning the analysis of our model's results must be clarified. 

First of all is the evaluation of the accuracy of saccades: several different criteria assessing the precision of in vivo saccades have been proposed in the litterature. Among them, \cite{McPeek2002} propose that accurate saccades land within a radius of up to one fifth of the target's eccentricity of the target's position, while \cite{McPeek2006} propose that accurate saccades land within a radius of $2\,^{\circ}$, or 15\% of the saccade's amplitude, of the target's position.

Secondly, we must note that all timings plotted from the model are counted from the appearance of the target in the Visual map of the SC. This does not take into account the latency between the presentation of a target in the visual field of an animal and the transmission of the retinal information to the superficial layers of the SC, which is estimated around 40ms for type I neurons (which respond well to the kind of stimuli usually used in experimental tasks, as measured by \cite{Rizzolatti1980}). 

Thus, these 40ms should be added to all timings given in the results below, when comparing the model's simulated events related to target appearance in the visual field with similar in-vivo experimental results. 

\subsection{Model characterization\label{part_3.1}}

\begin{figure}[h]
\includegraphics[scale=1.0]{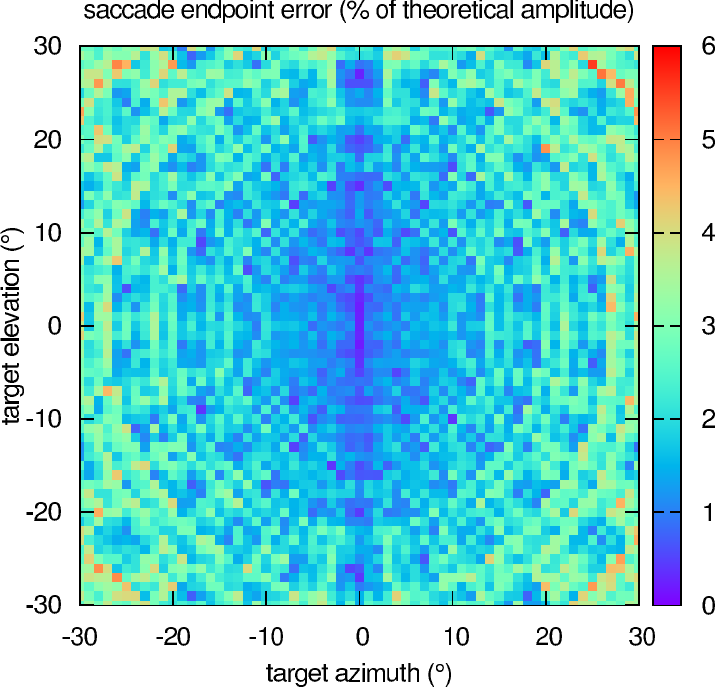}
\caption{\label{fig:carac_1C_accuracy}{\bf Saccades endpoint error map for Task 1.} the error is calculated as the ratio of the difference between the final eye position and the target position, and the amplitude of the desired saccade.}
\end{figure}

When only one target is displayed, the model is able to perform saccades with good accuracy across the whole range of the simulated visual field: the maximal error between the desired and obtained eye endpoints is inferior to 6\% of the desired amplitude of the saccade (see fig.~\ref{fig:carac_1C_accuracy}), well under the accuracy criteria evoked before.

\begin{figure*}[t]
\includegraphics[scale=1.4]{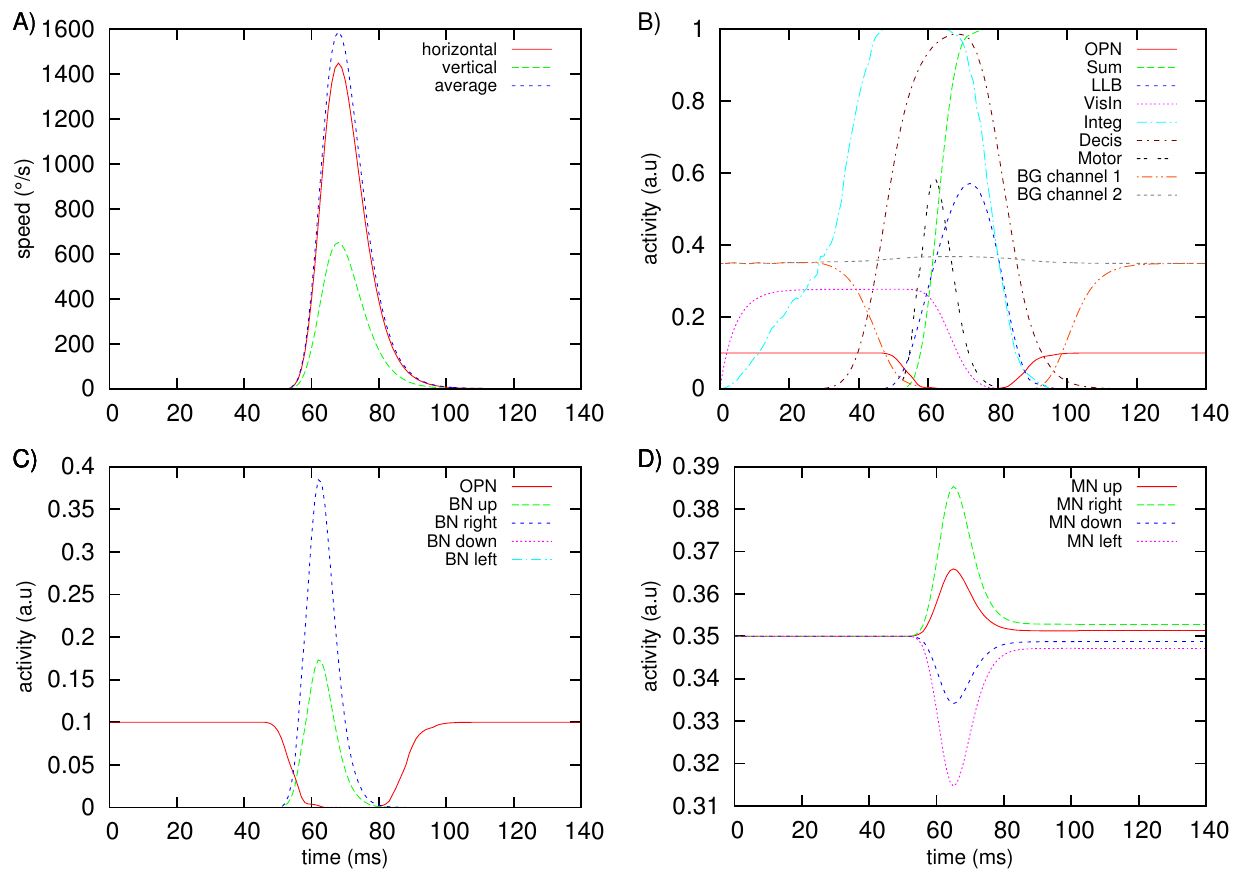}
\caption{\label{fig:carac_1C_SC-SBG profiles}{\bf Time course of SC maps activity with regards to target presentation.} Neurons activity profiles during a saccade towards a target located at coordinates $[20\,^{\circ},10\,^{\circ}]$, aligned with target presentation. A), eye horizontal, vertical and mean speed profiles. B), BG outputs for channels 1 (selected) and 2 (unselected), SBG OPNs and SC neurons coding for the center of the target's position in each SC map . C), 
OPNs and E-IBN for all four SBG circuits. D), MNs for all four SBG circuits. See text for abbreviations.}
\end{figure*}

Mean eye speed profiles (as examplified in fig.~\ref{fig:carac_1C_SC-SBG profiles}-A) show that saccade latency is close to 55ms after the onset of activity in the Visual map - that is 95ms after target presentation in the visual field, when retinal-to-SC input latency is taken into account, which is compatible with express saccade latencies (cf. \cite{Fischer1983}). 

The appearance of a target in the Visual map (fig.~\ref{fig:carac_1C_SC-SBG profiles}-B, purple curve) feeds a build-up integration by the Integration map (fig.~\ref{fig:carac_1C_SC-SBG profiles}-B, light blue curve). This Integration map activity is then fed to the BG, which results in a progressive desinhibition of the BG output selectively for the channels coding for the target's location (fig.~\ref{fig:carac_1C_SC-SBG profiles}-B, orange curve for the BG channel coding for the target's location,
vs grey curve for a channel coding for a section of SC integration map not stimulated by the target), starting around 30ms after the onset of Visual map activity. Consequently, the integration rate in the Integration map is increased specifically for the neurons coding the target's position, thus generating a burst of activity that speeds up the full disinhibition of the corresponding BG channel. Furthermore, the Integration map to Decision map connection is disinhibited specifically
for the neurons excited by the target, thus allowing the burst component of the Integration map activity to evoke a burst in the Decision map centered on the target's location (fig.~\ref{fig:carac_1C_SC-SBG profiles}-B, yellow curve). 

The parametrization of the LLB circuitry, a combination of high activation threshold and weak input weight, ensures that the LLB summation of this Decision map burst is quite slow (see fig.~\ref{fig:carac_1C_SC-SBG profiles}-B, dark blue curve), and explains the 20ms-delay between the onset of the Decision map burst and that of the Motor map burst (fig.~\ref{fig:carac_1C_SC-SBG profiles}-B, black curve).

The Motor map burst elicits activity in the E-IBNs of those circuits of the SBG, concerned with movement towards the position of the target, around 50ms after target presentation, gated by the OPNs inhibition (fig.~\ref{fig:carac_1C_SC-SBG profiles}-C, no BN activity is elicited for the leftward and downward circuits), which in turn generate a burst of MN activity for the corresponding SBG circuits (fig.~\ref{fig:carac_1C_SC-SBG profiles}-D, the MN of the up- and right-ward circuits are activated when those
of the left- and down-ward circuits are inhibited, in order to contract or relax the corresponding EOM) and produce the eye movement evoked earlier. The MNs will revert to a tonic activity, but on a different level than before target presentation, in order to keep the eyes fixed on their final position.

The Motor map burst also feeds the summation neuron responsible for the STT (fig.~\ref{fig:carac_1C_SC-SBG profiles}-B, green curve), which exert an inhibitory feedback on the Motor map, as well as on the Visual and Integration maps. The Motor burst has the shortest duration, and disappears around the same time the Visual map activity is completely shut down. The bursting part of the Integration map activity is sustained by the low leak of this map well after the Visual map extinction. As a consequence, the decision map burst is the last to be extinguished, causing the LLB inhibition of the OPNs to lift. 

\begin{figure}[h]
\includegraphics[scale=0.68]{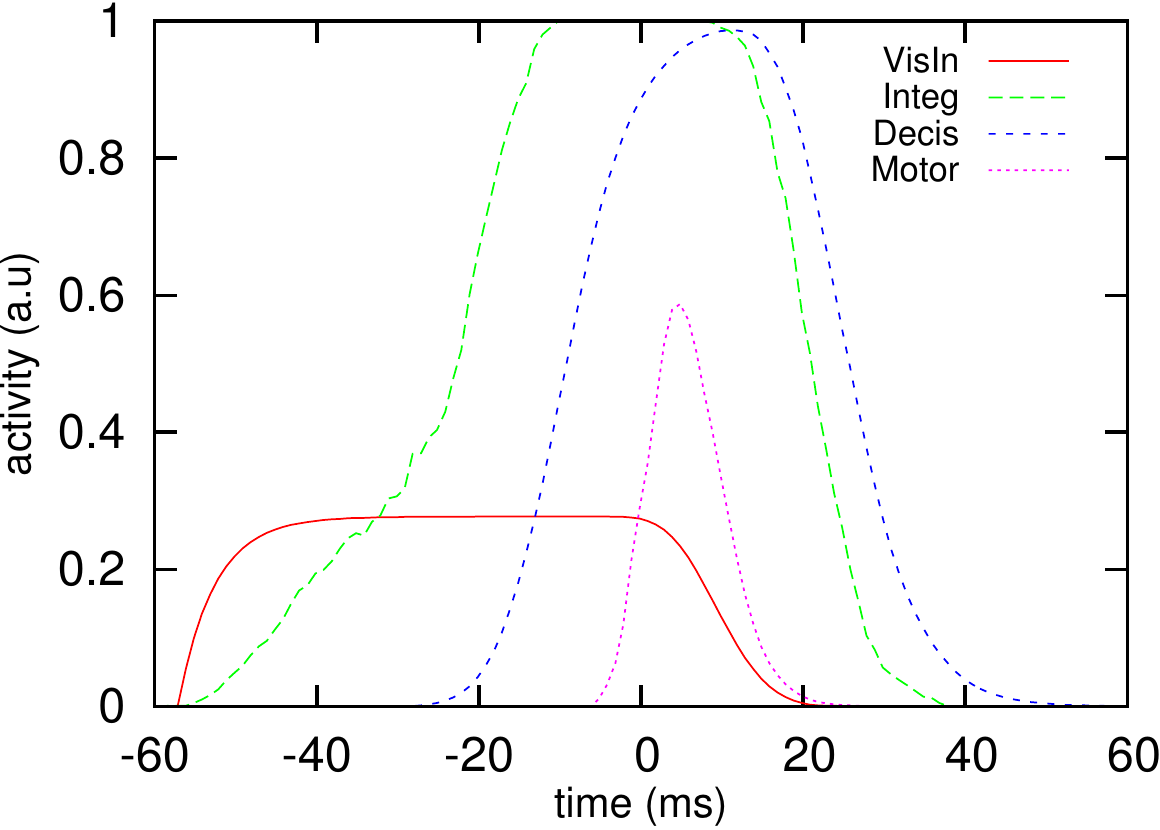} 
\caption{\label{fig:carac_1C_SC_comp-McPeek}{\bf Time course of SC maps activity with regards to saccade onset.} Neurons activity profiles for the Visual, Integration, Decision and Motor maps aligned with saccade onset, for a saccade towards a target located at coordinates $[20\,^{\circ},10\,^{\circ}]$.}
\end{figure}

Fig.~\ref{fig:carac_1C_SC_comp-McPeek} summarizes the time courses of the activity of the neuron coding for the target's position in the Visual, Integration, Decision and Motor maps when aligned to mean saccade onset: the precise timings of these activity profiles can be manipulated by parameter modifications (such as the LLB threshold, or the weight of the STT inhibitions), but their order is intrinsic to the model architecture.

These characteristics can be compared with the results shown in fig.~1 of \cite{McPeek2002}, that expose electrophysiological recordings of the motor-related activity of various neurons types : they identify several classes of SC neurons, especially the Visuo-Movement Burst Neurons (VBN), the Visuo-movement Prelude Neurons (VPN) and the Movement Neurons (Mvt.N). The motor-related activity of the VPNs shows a long build-up followed by a burst component centered on saccade onset. The VBN show no such build-up, but only a burst centered on saccade onset. Finally, the Movement Neurons' discharge pattern is also a burst centered on saccade onset, but with a narrower width than the burst of the VBNs. 

\begin{figure}[th]
\includegraphics[bb=25bp 15bp 300bp 250bp,scale=0.85]{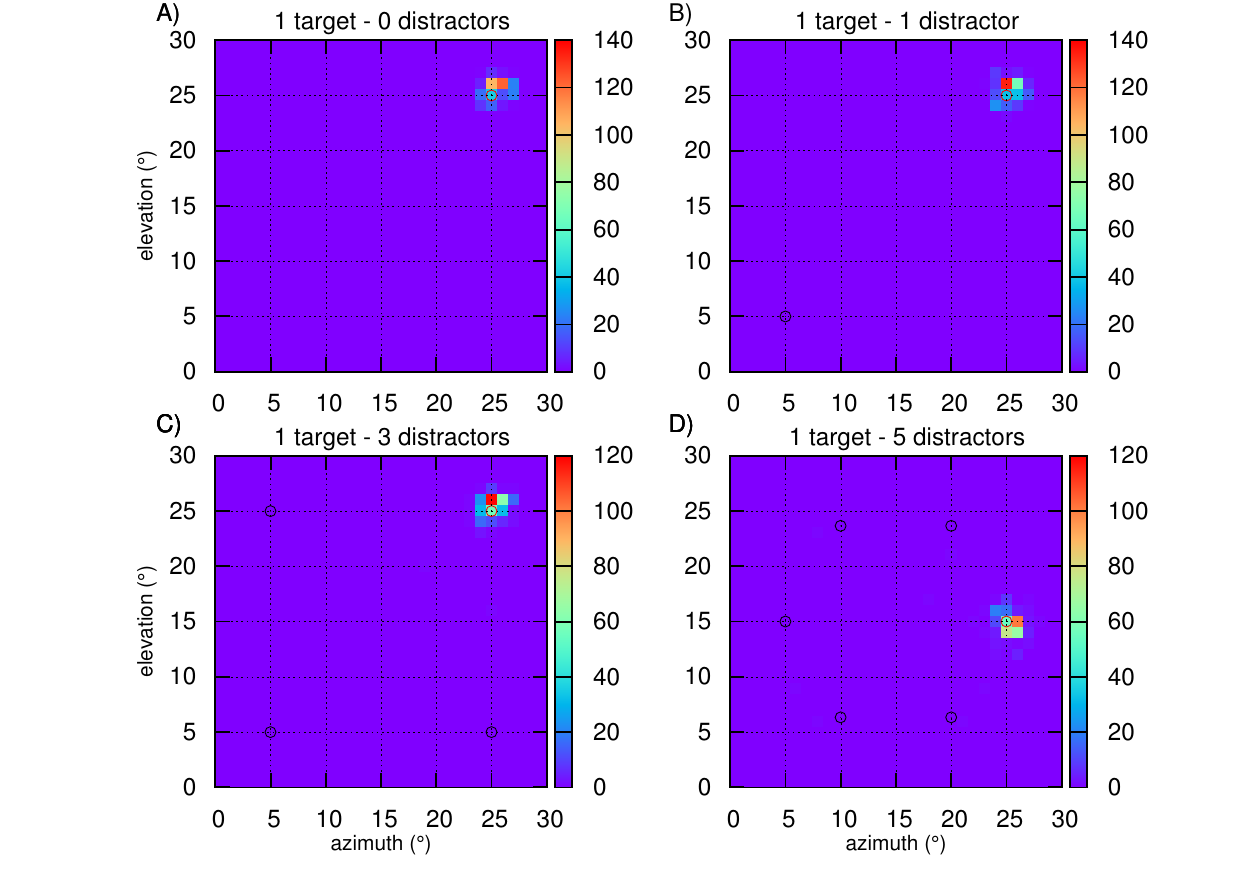}
\caption{\label{fig:distr_endpoints}{\bf Saccade endpoints distribution for the target-and-distractors setup of Task 2.} Endpoints distribution when the target is presented alone (top-left), with one distractors (top-right), with three distractors (bottom-left) or with five distractors (bottom-right), when target value is maximal. Red circle indicates target position, black circles distractors positions.} 
\end{figure}

Setting aside the burst of activity correlated with target presentation exhibited by the VPN and VBN, which has not been modelled in our work, the discharge patterns of the VPN is strongly similar to that of our Integration Map neurons activity profile, when the VBN are akin to our Decision Map neurons and the Mvt.N are akin to our Motor Map neurons. Therefore, we hypothesize that the role of the Integration Map, the Decision Map and the Motor map of our model is played in-vivo by the VPN, VBN and Mvt.N populations respectively. Our model is set so that the Motor map burst is temporally contained within the Decision map burst, which starts itself after the Integration map burst, a feature that the various experimental recordings of the VBN, VPN and Mvt.N by \cite{McPeek2002} do not allow to verify, but this prediction could nevertheless be tested in-vivo on these neuron populations, as well as all our model's assumptions and predictions concerning the connections and operations of the various SC maps.

\subsection{Selection between targets and distractors, or between distractors only\label{part_3.2}}

Fig.~\ref{fig:distr_endpoints} shows that the model is able to produce perfect selection in all conditions of the target-and-distractors setup, since no distractor is ever selected no matter their number, and saccade accuracy is as good as in task 1.

\begin{figure}[h]
\includegraphics[bb=8bp 15bp 300bp 250bp,scale=0.85]{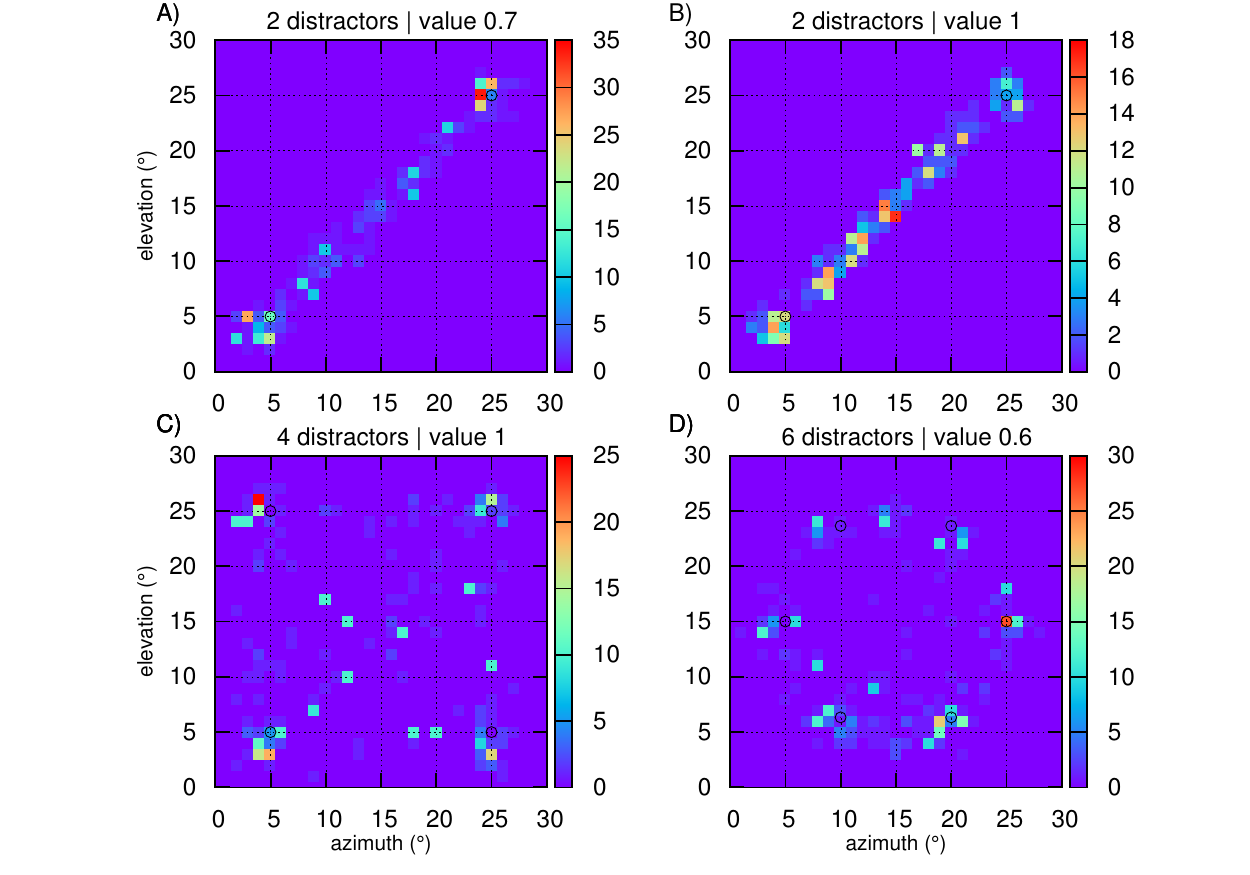}
\caption{\label{fig:targ_endpoints}{\bf Saccade endpoints distribution for the distractors-only setup of Task 2.} Endpoints distribution when two distractors of value 0.7 compete (top-left), when two distractors of maximal value compete (top-right), when four distractors  of maximal value compete (bottom-left) or when six distractors of value 0.6 compete (bottom-right). Black circles indicate distractors positions.}
\end{figure}

In the distractors-only setup, selection also occurs, and is accurate saccades made towards one stimulus (as opposed to average saccades made when multiple stimuli are selected) are stochastically distributed between all distractors, as experimentally shown in \cite{McPeek2004} (compare fig.~\ref{fig:targ_endpoints}-C with their fig.~2-e). When accurate saccade are made towards any distractor, saccade accuracy seems lower in this setup than in the target-and-distractors setup, since saccade endpoints are less tightly grouped around the chosen distractor (compared with the similar condition of the target-and-distractors setup).

It also appears that, in the distractors-only setup, selection is not always independant of the value of the distractors: when two distractors compete, the proportion of average saccades (which are a symptom of the simultaneous disinhibition of more than one competitor by the BG) is much higher when the distractors value is maximal (compare fig.~\ref{fig:targ_endpoints}-A for results with 2 distractors of value 0.7, and fig.~\ref{fig:targ_endpoints}-B for results with 2 distractors of maximal value).

\begin{figure}[h]
\includegraphics[scale=0.45]{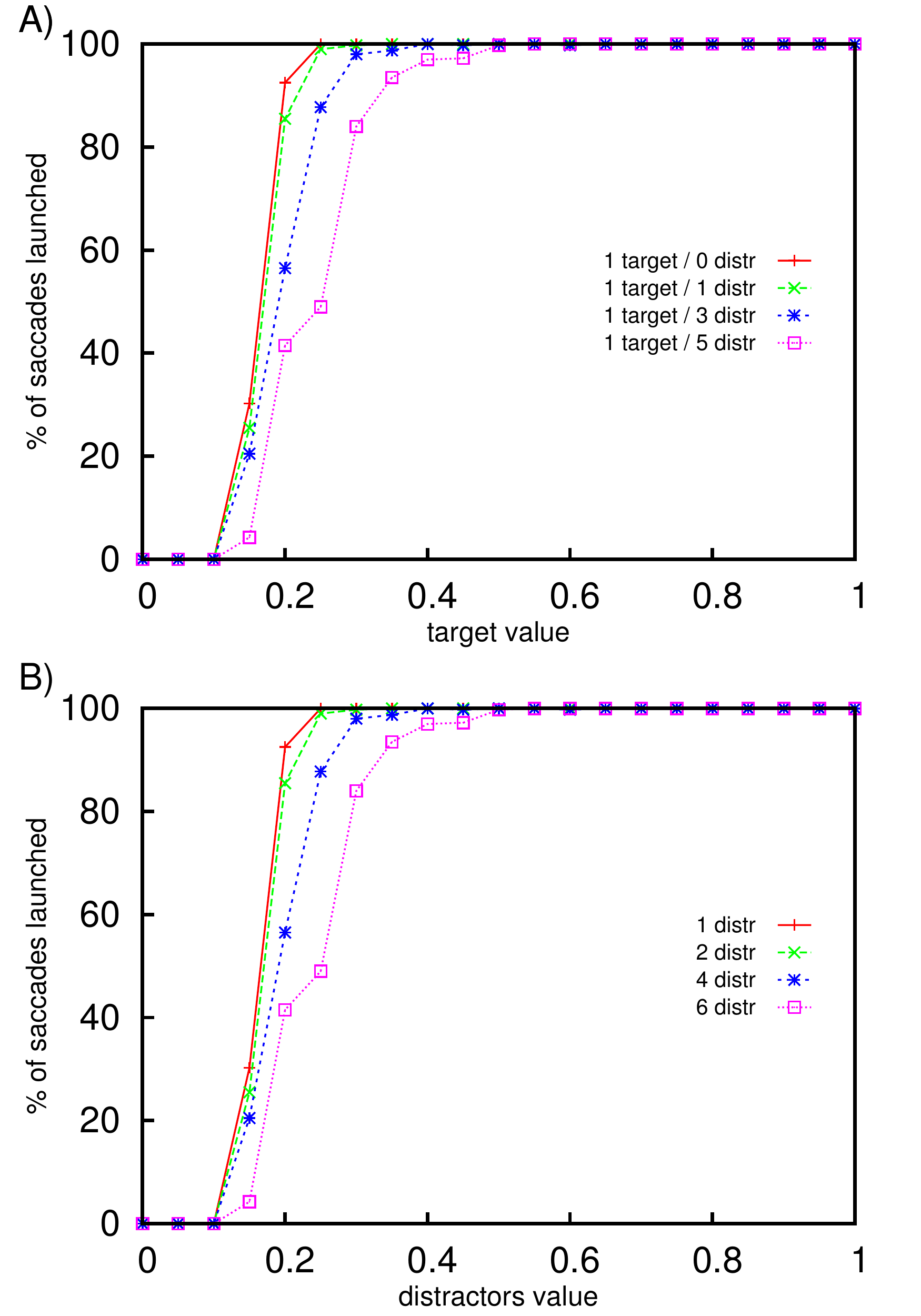}
\caption{\label{fig:distr_salience-initiation} {\bf Effects of target value on saccade initiation in Task 2.} A), in the target-and-distractors setup. B), in the distractors-only setup.}
\end{figure}

Furthermore, there is a threshold under which the value of a target is not sufficient to elicit a saccade (fig.~\ref{fig:distr_salience-initiation}-A, red curve). Interestingly, this threshold increases with the number of distractors presented simultaneously to the target (fig.~\ref{fig:distr_salience-initiation}-A), meaning that the more distractors compete with a target of intermediary value, the more difficult it is for the model to elicit saccades toward this target. This tendency remains moderate, and might thus be difficult to establish experimentally. 

The same tendency is also noticeable in the distractors-only setup: the value threshold for saccade elicitation increases with the number of competing distractors when 1, 2 or 4 of them compete (fig.~\ref{fig:distr_salience-initiation}-B, red, green and blue curves). Results in the six-distractors condition (fig.~\ref{fig:distr_salience-initiation}-B, pink curve) contradict this tendency: in this condition, the value threshold for saccade elicitation is lower than in the 4-distractors condition. 

In the 6-elements condition, the points competing in the visual field are close enough from each-other for the neurons of the Visual map stimulated by any given point of the visual field to be also stimulated by this point's neighbours. Thus, the effective value coded by the neuron coding for the center of the targets in the Visual Map is higher than their value coded by the corresponding neuron in the retinal input. This value increase between the retinal input and its Visual map representation is of course reflected over the whole 2D-Gaussian representation of each competing target, and is proportionnaly bigger for the neurons at the periphery of the Gaussian than for the neuron at its center, thus greatly increasing the total value encoded by said Gaussian.

In the target-and-distractors setup, the effective value in the Visual map of the element of the visual field which will be selected - here the target - is around 2\% higher for the center of any point's representation in the 1 target and 5 distractors condition (the 6-elements conditions of this setup) than in all other conditions, in the $[20,50]ms$ interval after target onset where competition is always ongoing whatever the saccade latency. The effective value in the same map for the element which will be selected in the distractors-only setup - that is any distractor - is around 4\% higher for the center of any point's representation in the 6-distractors conditions than in the any other condition (1, 2 and 4 distractors), in this same timeframe.

\begin{figure*}[t]
\includegraphics[scale=1.3]{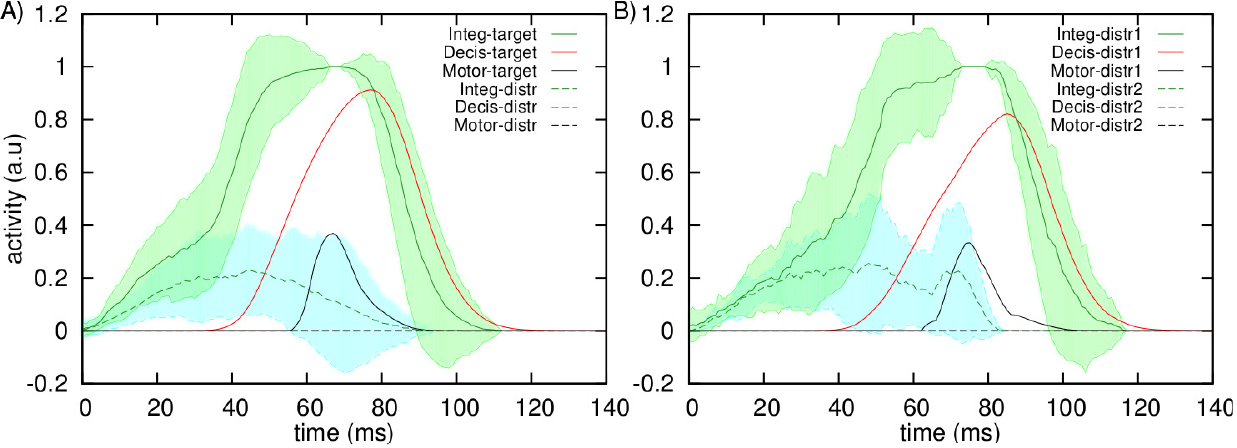}
\caption{\label{fig:comp_VisInput_task2} {\bf Variation of the activity profiles in the SC maps with regards to the difficulty of the selection.} Activity profiles of Visual (blue), Integration (green), Decision (red) and Motor (black) map neurons coding for the positions of two competing points in the visual field. A), averaged over 500 trials in the one target (maximal value) and one distractor condition of the target-and-distractors setup. B), averaged over the 130 trials where only distractor 1 was selected in the two distractors (maximal value) condition of the distractors-only setup. The green area represents the standard deviation around the mean activity of the integration neuron of the winner of the competition, and the blue area the same for the activity of the integration neuron of the loser of the competition.}
\end{figure*}

This is similar to each distractor of the 6-elements condition of the distractors-only setup having more value than expected, hence the discrepancy in value threshold for saccade elicitation noted for this condition in fig.~\ref{fig:distr_salience initiation}, bottom. 

In the target vs. distractors protocol, the VPNs recorded by \cite{McPeek2002} exhibit a ramping activity followed by either a burst (target case) or a gradual decrease of activity (distractor case). They show (their Fig. 10) that in these two different cases, the ramps are at first undistinguishable, and become distinguishable only after integration has been going on for a while. In our simulations, the neurons of the integration map also have a ramping activity (fig.~\ref{fig:comp_VisInput_task2}-A, continuous and dashed green traces), followed by either a decrease (distractor, dashed trace) or a burst (target, continuous trace). During this ramping phase, the average slope depends on whether the neuron codes for a distractor or for the target. However, because of noise, the variability of these average firing rates is so large (Fig.~\ref{fig:comp_VisInput_task2}-A, light blue and light green areas represent the standard deviation around these means), that the divergence of these ramping activities can be only assessed late in the integration process. We thus assume that our Integration map neurons are a model for the VPNs recorded in \cite{McPeek2002}.

In the same protocol, our Decision neurons (Fig.~\ref{fig:comp_VisInput_task2}-A, red traces) can be likened to the activity profiles of the Visuo-Motor Burst neurons (VBN) lacking a strong second visual burst (Fig. 7 of \cite{McPeek2002}): in both cases, the neurons have no activity until selection is reached, and only the neuron coding for the selected target exhibit a burst centered on saccade onset. 

The model's Motor map neurons exhibit the same pattern of activity when coding for either a selected or unselected target as the Movement neurons recorded in fig.~13 of \cite{McPeek2002}, emitting a burst only when the selected target is in their receptive field (fig.~\ref{fig:comp_VisInput_task2}-A, black curves). 

Finally, the activity profiles obtained for the Integration map neurons in the simulation of the distractors-only task (Fig.~\ref{fig:comp_VisInput_task2}-B, continuous and dashed green traces) predict that in such a case, the discriminability between a distractor about to be selected for a saccade and another distractor should be delayed even longer (see in fig.~\ref{fig:comp_VisInput_task2}-B the overlap in the green and blue area representing the variability of the integration for each distractor), up to the beginning of the burst associated to movement itself. After the identification of VPNs, this condition could easily be tested experimentally.

\begin{figure*}[!]
%\center\includegraphics[bb=50bp 10bp 300bp 200bp]{fig11/figure_11.pdf}
\center\includegraphics[scale=1.0]{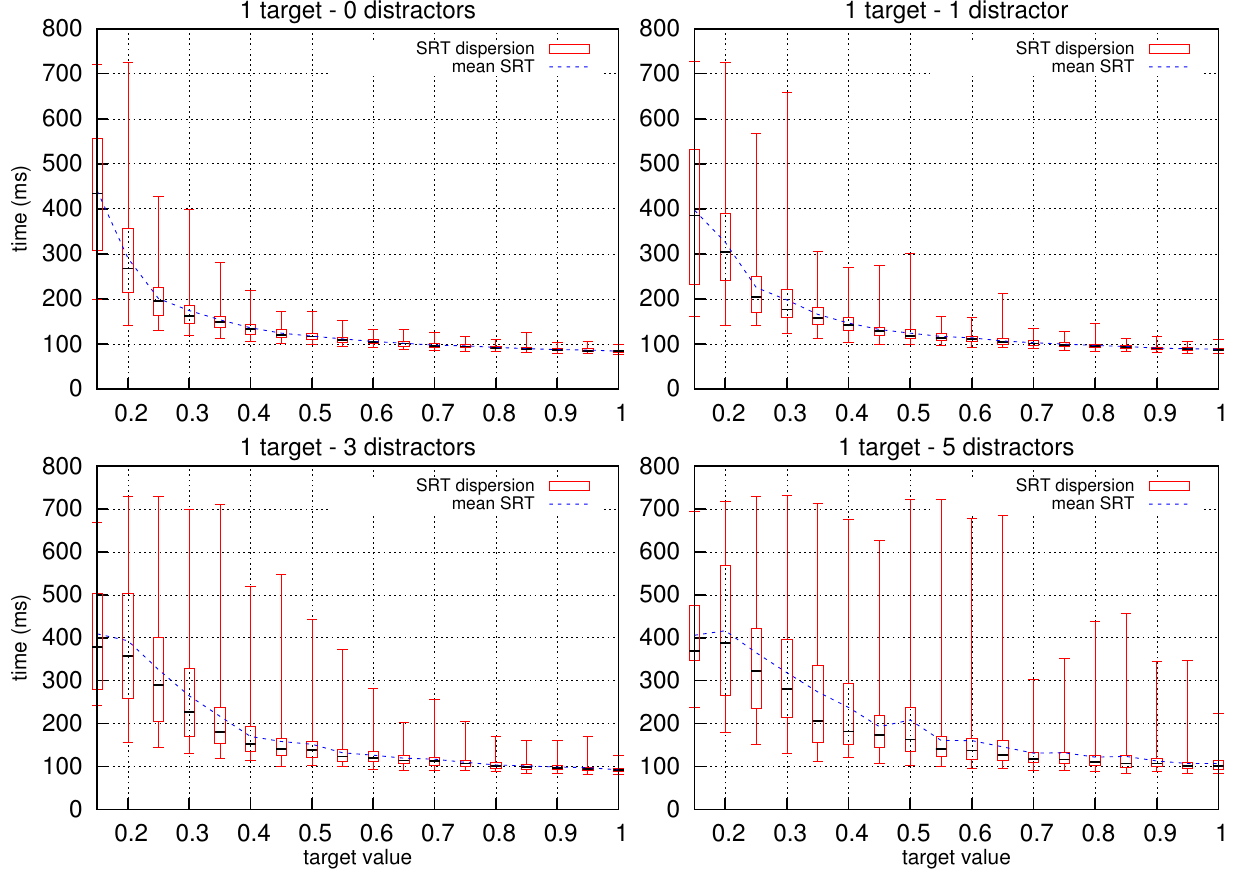}
\caption{\label{fig:distr_salience-latence}{\bf Effects of target value on saccade latency in the target-and-distractors setup of Task 2}. Mean latency and latency dispersion with regards to target value when the target is presented alone (top-left), with one distractors (top-right), with three distractors (bottom-left) or with five distractors (bottom-right)}
\end{figure*}

\begin{figure*}[!]
%\center\includegraphics[bb=50bp 10bp 300bp 210bp]{fig12/figure_12.pdf}
\center\includegraphics[scale=1.0]{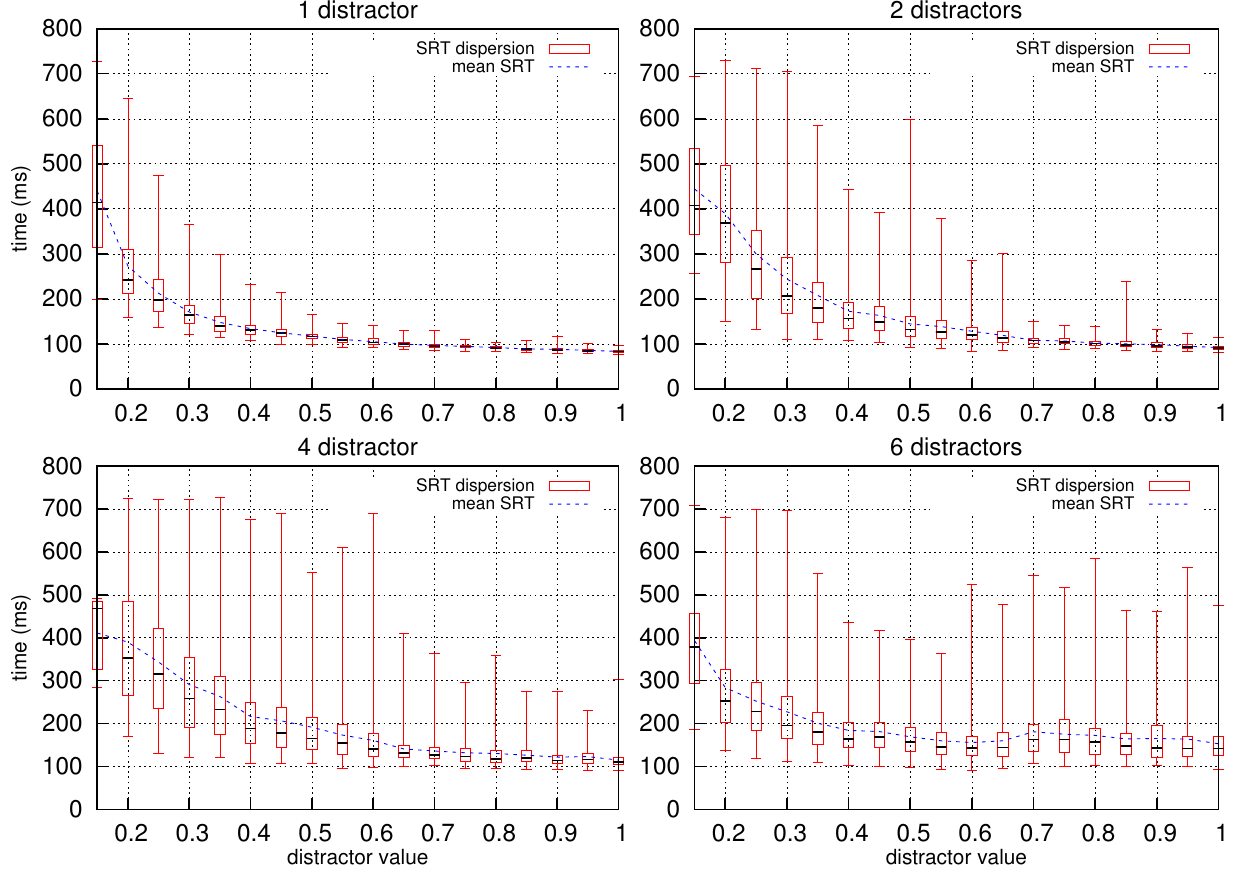}
\caption{\label{fig:targ_salience-latence}{\bf Effects of distractors value on saccade latency in the distractors-only setup of Task 2.} Mean latency and latency dispersion with regards to distractor value when only one distractor is presented (top-left), two distractors (top-right), four distractors (bottom-left) or six distractors (bottom-right)}
\end{figure*}

Concerning saccade latency in this task (that is the delay between the presentation of an element in the visual field and the beginning of the eye movement - also called Saccadic Reaction Time, or SRT), our model presents two results: firstly, the SRT decreases when the value of the elements in the visual field increases (see the decrease of mean latencies as well as latency dispersions in each of the four panels of fig.~\ref{fig:distr_salience-latence} for the target-and-distractors setup, and the same in fig.~\ref{fig:targ_salience-latence} for the distractors-only setup). This is consistent with experimental results in tasks where the discriminability between the stimuli and the background of the visual panel is low. 

\begin{figure}[h]
\includegraphics[scale=0.5]{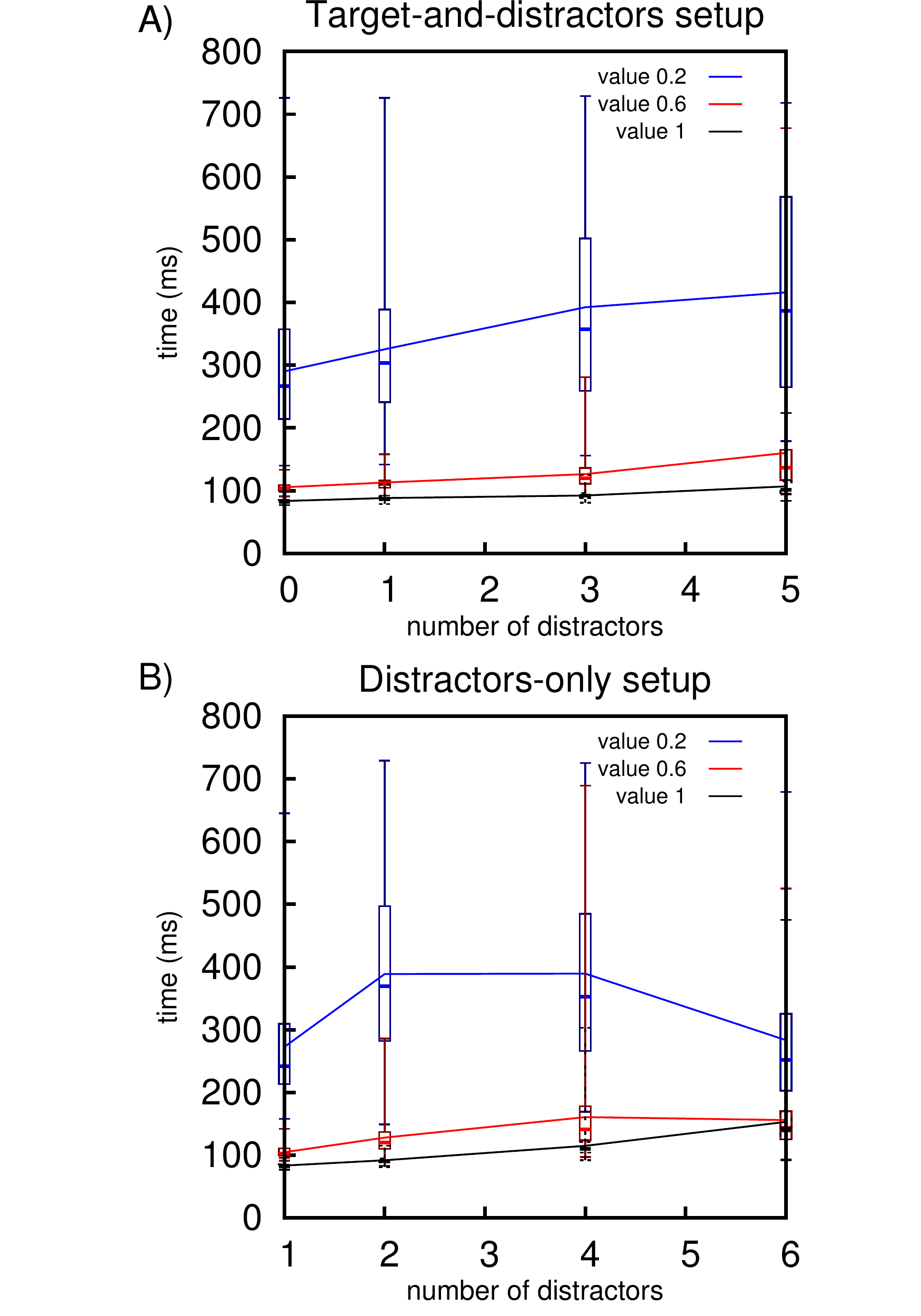}
\caption{\label{fig:Task4_RDE}{\bf Evolution of saccade latencies with the number of distractors and the value of the target in Task 2.} A), target-and-distractors setup. B), distractors-only setup.} 
\end{figure}

Secondly, the SRT increases with the total number of elements displayed in the visual field. Mean latency as well as latency dispersion around the median value increase for a given value with the number of points in competition, whether they be target and distractors or distractors only: see the comparison of the four panels of fig.~\ref{fig:distr_salience-latence} to each-other, as well as those of fig.~\ref{fig:targ_salience-latence}, which is resumed in fig.~\ref{fig:Task4_RDE}.

This phenomenon, called Remote Distractor Effect or RDE, seems less obvious in the distractors-only setup than in the target-and distractors setup when the distractor value is low (compare fig.~\ref{fig:Task4_RDE}-A and bottom, blue curves), but saccade latency is markedly higher in both setups for any given value when the number of competing points increases (red and black curves). In this case, the RDE is more pronounced for the distractors-only setup than for the target-and-distractors setup: the increase in saccade latency is much larger when going from 1 to 6 distractors of maximal value (fig.~\ref{fig:Task4_RDE}-B, black curve), than when going from 1 target to 1 target and 5 distractors (fig.~\ref{fig:Task4_RDE}-A, black curve).

In our case, the source of the RDE lies in the diffuse projections within the basal ganglia loop, which are the substrate of competition between options. Indeed, while each channel is trying to disinhibit its output in the GPi through the Striatum to GPi focused inhibitions, it also tries to inhibit its neighbours through the diffuse inhibitions of the TRN over the Th and of the FS over the Striatum, as well as through diffuse excitations of the STN over the GPi. As a consequence, when the number of competitors is increased, the effects of this competition is stronger: each competitor, even the one with the highest input, sees its representation in the TRN and the Striatum attenuated by the additional inhibitions. 

This diminishes the Striatum to GPi self-inhibition, which is the source of the selection, while it additionally receives more activity from its neighbors in the GPi from the STN, where the GPe inhibitory feedback is not sufficient to compensate for this increase. In the end, the resulting attenuation of the potential winner delays the moment when selection is complete.

\begin{table}[h]
\begin{tabular}{|c|c|c|c|}
\hline 
setup/condition & 1T - 1D & 1T-3D & 1T-5D\tabularnewline
\hline 
target - distractors & +38\% & +58\% & +86\%\tabularnewline
\hline 
 & 2D & 4D & 6D\tabularnewline
\hline 
distractors & +80\% & +105\% & +123\%\tabularnewline
\hline 
\end{tabular}
\caption{\label{tab:RDE_proportions} {\bf Total GPi input increase in conditions with multiple distractors, when compared to the 1 target or 1 distractor only reference.} T: target, D: distractor. Values calculated over at least 400 trials where accurate saccades have been made towards the target for each condition of the target-and-distractors setup, and over at least 200 trials whre accurate saccades have been made towards any distractor in the distractors-only setup}
\end{table}

We measured this attenuation by measuring the total inputs to the GPi in the $[10,40]ms$ interval after target presentation, where we are sure that the competition is still running, even in the 1 target case. The increased activity in the GPi, caused by the additional inputs from competitors, and causing the RDE, is quite strong when one target of maximal value is confronted to an increasing number of distractors (the distractors have half of the taget value, Table~\ref{tab:RDE_proportions}, first line), and even stronger when only distractors of maximal value are present (i.e. all inputs have the same value, Table~\ref{tab:RDE_proportions}, second line). 

\subsection{Effects of target separation on the production of accurate saccades\label{part_3.3}}

\begin{figure}[h] 
\includegraphics[scale=0.5]{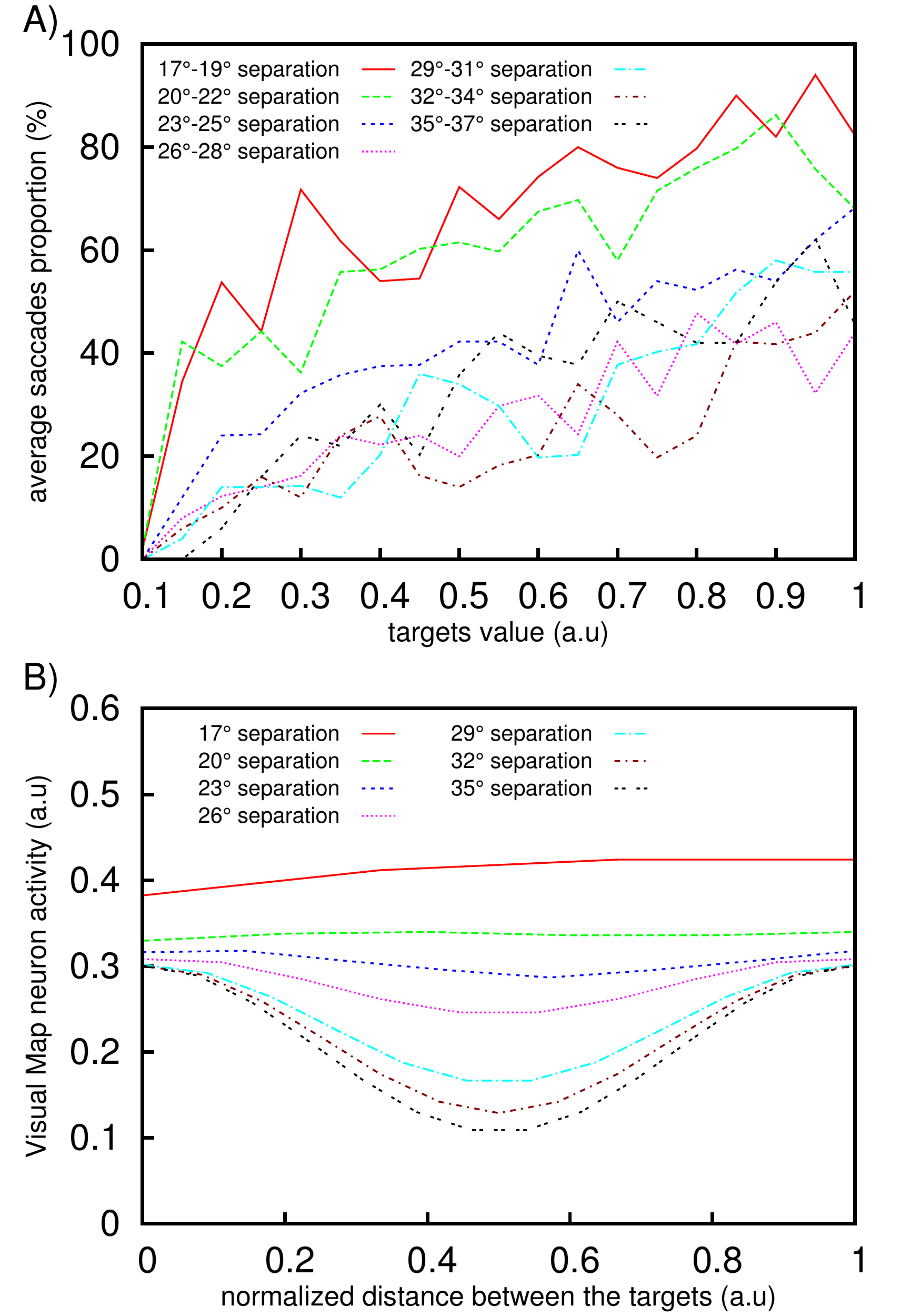}
\caption{\label{fig:metric_2C_proportion-average}{\bf Effects of target separation in Task 3 on the production of saccade and the overlap of activity loci in the Visual map of the SC.}A), proportion of average saccade with regards to targets value, averaged over separation intervals of $3\,^{\circ}$ for ease of representation. B), peak activity for all Visual map neurons located directly between the positions of the competing targets of value 1, for the same target separations intervals sampled for graph A).}
\end{figure}

For any given distance separating the targets, the proportion of average saccades increases with the value of the targets (see fig.~\ref{fig:metric_2C_proportion-average}-A, all curves). Furthermore, this proportion seems also dependant with target separation: the proportion of average saccades is bigger for distances under the $20\,^{\circ}$ threshold (fig.~\ref{fig:metric_2C_proportion-average}-A, red and green curves respectively representing the proportion of average saccade averaged over the $17$ to $19\,^{\circ}$ separation interval, and the $20$ to $22\,^{\circ}$ separation interval) than for separations greater than $23\,^{\circ}$ (fig.~\ref{fig:metric_2C_proportion-average}-A, other curves). 

Fig.~\ref{fig:metric_2C_proportion-average}-B, reveals that the Visual map neurons located directly between the two targets positions are more stimulated for small target separations, since in these cases the gaussians generated by each target significantly overlap.

When target separation is smaller than $20\,^{\circ}$, this overlap generates enough activity between the targets representations in the Visual map to seriously compete with the targets for selection, and thus regularly causes average saccades. Targets  separated by a greater distance suffer less from this effect, and only when they have high values.

This result can be compared with the effects of the quantity of competing elements in the visual field in Task 2, in which we saw that the representation of each element in the Visual map can receive additionnal stimulation because of the proximity of its neighbours. The coordinates given in Table~\ref{tab_positions-McPeek} allow to calculate the separation between all points in each conditions: $56.6\,^{\circ}$ for the 2-elements condition, $40\,^{\circ}$ for the 4-elements condition, and $20\,^{\circ}$ for the 6-elements condition. This $20\,^{\circ}$ separation is precisely the threshold after which two neighbouring points significantly overlap each-other, with the effects seen earlier.

In our model, the precise value of this distance threshold is dictated by parameter $\sigma$, which governs the size of the target's representation in the Visual map. Thus, this threshold can be adjusted by the tuning of $\sigma$. However the prediction of the existence of such a threshold is inherent to the model structure.

\section{Discussion\label{part_4}}

In this work, we proposed a biologically constrained model of the SC-BG loops that accounts for saccade target selection by way of the race-like competition of visual targets representations in the retinotopic maps of the superficial layers of the SC ; the BG themselves act as a threshold detector for the evidence accumulated by each target during competition. 

We showed that the activity profiles of the neurons of the various maps modelled in our SC are comparable to in-vivo recordings in the Monkey SC performed by \cite{McPeek2002}, and we therefore proposed a role for the Visuo-Motor Prelude and Burst neurons of the intermediary layers of the SC. We also propose a novel interpretation for the colliculu-baso-collicular loops based on convergent afferences from the SC maps to the BG channels, on divergent symetrical inhibitory efferences from the BG channels to the projection between the SC superficial and intermediary maps to one another, and on different roles for these efferences, one modulating connections weights to amplifiy contrasts between stimuli, and the other gating the transmission of signal to the deeper layers of the SC only for the stimulus having won the competition. Furthermore, we were able to reproduce and propose an explanation for specific experimental data gathered by \cite{McPeek2004} regarding the effects of local SC inactivation on the selection between one target and multiple distractors.

We predict a specific order and general shape for the activity in the SC maps in various selection tasks that can be tested in vivo; we also predict that the occurence of average saccades is linked to both the separation between the points in competition in the visual field, as well as the value of their representations in the SC Visual map. The quantitative aspects of these predictions can be adjusted by the fine tuning of the model's parameters in order to fit optimally with experimental data, but these predictions are qualitatively intrinsic to our model's architecture.

Furthemore, we propose a mechanism for the well-known Remote Distractor Effect, based on balance of diffuse and specific excitations and inhibitions exerted on each channels of BG's output nuclei, that increase the level of inhibition the winning target must lift on itself when more points compete in the Visual field.

We also predict that the coupling of the BG characteristics with the properties of the SC mapping can turn the RDE into a facilitating effect on saccade initiation in the specific case of the competition between an increasing number of similar stimuli, when the separation between stimuli reaches a critical minimal threshold.

\bigskip{}

The saccades produced towards single targets by our model are accurate, as exposed in section \ref{part_3.1}, and this accuracy is maintained when multiple stimuli of distinct value compete (see the results of the target-and-distractors setup in section \ref{part_3.2}). Yet, saccade accuracy degrades when multiples stimuli of similar value compete, with more average saccades being produced, and accurate saccades being more dispersed around their chosen stimulus. The absence of a Cerebellum module in the model could link this to the observation that cerebellar inactivation leads to a degradation of saccade accuracy (\cite{Iwamoto2002}). The addition of a Cerebellum module to the model, that would receive an efferent copy of the SC output to the SBG and correct it on-the-fly would increase saccade accuracy by allowing minute modifications of saccade trajectory when the saccade endpoint proposed by the SC is outside the confidence interval of the selected stimulus by a defined margin (which would have to be tuned so as to still allow for average saccades, and not suppress them completely).

Another feature lacking in the model is the implementation of a mechanism modeling the variation of saccade duration with target eccentricity, in order to respect the parameters of the main sequence of in-vivo saccades (as defined by \cite{Bahill1975a}). Such a mechanism could be based on the proposals of \cite{Goossens2006,Opstal2008}, which modulates the maximal activity of neurons in the SC motor map according to the metrics of the saccade they code.

\bigskip{}

Several models of selection use a principle of stochastic gated accumulators very similar to our model (\cite {Purcell2010,Schall2011,Purcell2012} amongst others) applied in the context of the cortical loops of the saccadic system, but do not model the source of the thresholds they use to gate the selection process. Nevertheless, they propose various operators for this gating mechanism, amongst which the BG. Our model developps this proposal by providing a biologically compatible loop between the SC and its evidence integrators, and the BG which inhibitory output dynamically reacts with the accumulation of evidence in order to gate the winning and losing stimuli. The use of BG output in order to gate the SC activity for the selection was already proposed by other SC models, such as \cite{Arai2004,Arai2005}, but these models used only a static gating output from the SNr to the SC maps which, contrary to our model, is not produced by a real-time disinhibition process in the BG fed by the SC maps, and therefore is not able to take into account the effects of any visual input variation on selection.

\smallskip{}

Furthermore, both cortical (\cite{Purcell2010,Schall2011,Purcell2012}) and subcortical (\cite{Arai2004,Arai2005}, but also the stochastic accumulator model of the SC by \cite{Ludwig2007}) models of saccade target selection implement the minimal amout of neurons needed to obtain selection, often in the form of two layers (or even just two sets of neurons, with only one neuron for each competing stimulus in each set), a visual one serving as an input-receiver that operates the race to selection, and a motor one where the movement command is generated. Our model propose a more accurate (although not complete, as exposed before) representation of the anatomical constaints of the biological structures involved in selection (the SC and its connections to the BG), with several new loops and intermediate layers in the SC for which we propose roles and biological substrates not previously accounted for, and with a biologically plausible generation of the motor command by a dedicated structure (the SBG).

\smallskip{}

Lastly, the cortical models of saccade target selection can only account for some of the saccade timings observed in-vivo: the latency of the signal processed through the visual cortex to the FEF being on par with the latency of the express saccades, the selection processes for these saccades cannot be explained by cortical models. The saccades latencies generated by our model range from $90$ to $100ms$ in cases where no selection or an easy selection is made, to more than $200ms$ in cases where difficult selection must be resolved before initiating movement. This first range of simulated latencies is consistent with the express saccades observed in-vivo (\cite{Fischer1983}), even though the model does not feature the fixation cells in the rostral SC commonly associated with the production of express saccades \cite{Munoz1992,Munoz1993,Munoz1993a}. The second range of simulated latencies reaches values where one would expect the intervention of the cortical actors of the saccadic system - most notably the FEF, which are not modelled here, and therefore bridge the gap between purely subcortical and purely cortical models.  

This addition could modifiy the results of selection by modulating the values of the various stimuli represented in the SC maps (for example favouring the selection of a color over the location of a specific set of coordinates in the visual field), features and problems partially addressed under the more specific light of reinforcement learning by the model of \cite{NGuyen_a} 

\section*{Acknowledgments}

This work was partially funded by the HABOT project from the Emergence(s) program of the Ville de Paris (France).

\appendix

\section{Calculation of the Retinal input to the Visual map and procedure of gluing\label{sec:retinal_input}}

\begin{figure}[h]
%\center\includegraphics[bb=0bp 0bp 908bp 1119bp,scale=0.25]{fig-annexe1/figure_annexe_1.pdf}
\center\includegraphics[scale=0.25]{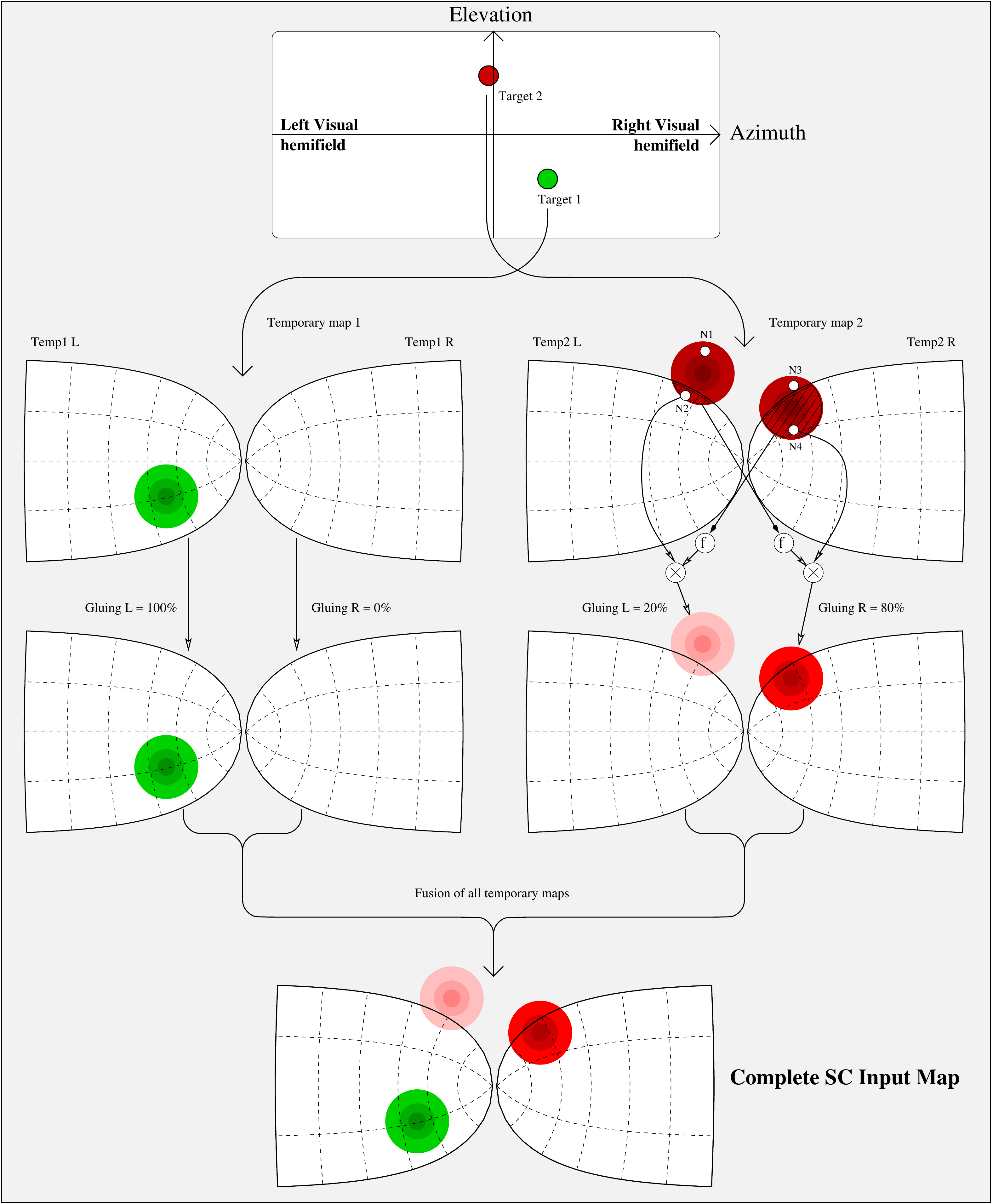}
\caption{\label{fig:annexe-1_SC-Gluing}{\bf Gluing scheme for the projection of the Retinal input to the Visual map.} The green stimulus is far enough from the vertical axis separating the two colliculi to be represented on only one collicular map; the red stimulus is close enough from the vertical axis to be represented on both collicular maps, so each of these representations has to be weighted to give both green and red targets the same total activity. The total activity generated within the boundaries of the Visual field in the left collicular map (hatched colored area in map Temp2L, containing neuron N2) is inversely modulated by the proportion of the total activity generated in the contralateral map within the boundaries of the Visual field (hatched colored area in map Temp2R), and vice-versa for the activity generated in the right collicular map. Therefore, the complete Visual input to the SC has a stronger component for the red target in the right colliculus than in the left colliculus}
\end{figure}

\indent

Wathever the specific conditions of any given task protocol, the retinal input fed to the SC Visual Map is calculated as such:
\begin{enumerate}
\item Each point $P{}_{i}$ presented in the visual field is characterized by its position in azimuth and elevation $[az_{i},el_{i} $ and its value $val$$_{i}$. Its position is translated into coordinates that will represent the position $[x_{i},y_{i}]$ in the SC maps as per equation (2). Fig.~\ref{fig:annexe-1_SC-Gluing} features two targets, one near the vertical axis (red dot) and one away from it (green dot). 
\item A temporary retinotopic map $temp_{i}$ is created for each point $P{}_{i}$ , which activity is represented by a 2D-Gaussian with standard deviation $\sigma=2.5$ neurons ($0.35mm$ in a Monkey SC), centered on the neuron at coordinates $[x_{i},y_{i}]$ in our discretized maps , and of maximal height equal to $sal_{i}$. 
\begin{enumerate}
\item If $P{}_{i}$ is far enough from the vertical axis to be represented by a Gaussian fully located in one of the colliculi (the one contralateral to the visual hemifield where $P{}_{i}$ lies), nothing special occurs (cf. the green target in fig.~\ref{fig:annexe-1_SC-Gluing}, which is fully located in the right visual hemifield, and whose representation in the retinal input is restricted to the left SC). 
\item If $P{}_{i}$ is close enough to vertical axis (cf. the red target in fig.~\ref{fig:annexe-1_SC-Gluing}), its Gaussian representation may cross the boundaries of the SC contralateral to the visual hemifield where $P{}_{i}$ lies (cf. red target representation in temporary map Temp2 R in fig.\ref{fig:annexe-1_SC-Gluing}). In such cases, another gaussian is calculated in order to represent the portion of the activity generated by this target in the other SC (cf. red target representation in temporary map Temp2 L in fig.~\ref{fig:annexe-1_SC-Gluing}). These two Gaussians have to be weighted down in order to keep the total activity generated by the target constant, whether it is represented by one gaussian in one colliculus, or two gaussians in the two colliculi. Thus, the maximum level of activity of each gaussian is modulated by a gluing factor calculated by the transfer function described in equation \ref{eq:fSigmo}. The gluing for each Gaussian is linked to the difference between the summed activity generated by the target within the boundaries of the colliculus contralateral to the Gaussian in question, and the summed activity generated by the target within the colliculus ipsilateral to the Gaussian in question. This solution is similar to the motor gluing proposed in \cite{Tabareau2007}.
\end{enumerate}
\item Once temporary maps have been calculated for all the points displayed in the visual field, all temporary maps are then merged in order to produce one single and complete map of all targets, that will serve as the input to the SC Visual map.
\end{enumerate} 
\smallskip{}

\begin{eqnarray}
Gluing_{ipsi} & = & 1-\label{eq:fSigmo}\\
 &  & \frac{1}{1+\exp(0.5\times min(\varphi,(I_{contra}^{tot}-I_{ipsi}^{tot}))}\nonumber 
\end{eqnarray}

With $Gluing_{ipsi}$ the gluing factor to be applied to the portion $I_{ipsi}^{tot}$ of the representation of any point of the Visual field specific to the ipsilateral colliculus, $I_{contra}^{tot}$ the equivalent of $I_{ipsi}^{tot}$ for the contralateral colliculus, and $\varphi$ a minimal threshold for the difference between the two summed activities.

\section{Model parameters\label{sec:tables_parameters}}

\begin{table}[h]
\center%
\begin{tabular}{|cc|cc|cc|}
\hline 
$NbCell$ & $43$ & $\omega_{Int}^{Dec}$ & $1$ & $Noise$ & $0.22$\tabularnewline
$NbChan$ & $11$ & $\omega_{Dec}^{LLB}$ & $0.03$ & $T_{BG}^{Int}$ & $0.359$\tabularnewline
$A$ & $\pi/60$ & $\omega_{Dec}^{Mot}$ & $1$ & $T_{BG}^{Dec}$ & $0.349$\tabularnewline
$B$ & $1.5$ & $\omega_{OPN}^{Mot}$ & $40$ & $\varepsilon$ & $10^{-10}$\tabularnewline
$\tau$ & $5ms$ & $\omega_{Mot}^{Sum}$ & $0.005$ & $Max$ & $0.3$\tabularnewline
$\tau_{small}$ & $3ms$ & $\omega_{Sum}^{Vis}$ & $0.1$ & $\varphi$ & $700$\tabularnewline
$L$ & $0.05$ & $\omega_{Sum}^{Int}$ & $0.35$ & $\sigma$ & $2.5$\tabularnewline
$G$ & $1$ & $\omega_{Sum}^{Mot}$ & $1$ &  & \tabularnewline
$\omega_{Vis}^{Int}$ & $11.2$ & $E_{LLB}$ & $0.5$ &  & \tabularnewline
\hline 
\end{tabular}
\caption{\label{tab_param_SC}{\bf Parameters of the SC model}}
\end{table}

\begin{table}[h]
\center%
\begin{tabular}{|cc|cc|cc|}
\hline 
$E_{OPN}$ & $0.1$ & $\omega_{BN}^{MN}$ & $0.101$ & $\omega_{MN}^{\theta_{acc}}$ & $16000$\tabularnewline
$\omega_{OPN}^{BN}$ & $20$ & $\omega_{BN}^{TN}$ & $0.0021$ & $\omega_{\theta_{vit}}^{\theta_{acc}}$ & $0.6$\tabularnewline
$\omega_{LLB}^{OPN}$ & $2$ & $E_{TN}$ & $0.35$ & $\omega_{\theta_{pos}}^{\theta_{acc}}$ & $4$\tabularnewline
$\tau_{small}$ & $3ms$ & $\omega_{TN}^{MN}$ & $1$ &  & \tabularnewline
\hline 
\end{tabular}
\caption{\label{tab_param_SBG} {\bf Parameters of the SBG model}}
\end{table}

\begin{table}[h]
\center%
\begin{tabular}{|cc|cc|cc|}
\hline 
$\tau$ & $5ms$ & $\omega_{Th}^{TRN}$ & $0.2$ & $\omega_{D1}^{GPe}$ & $0.5$\tabularnewline
$\tau_{small}$ & $3ms$ & $\omega_{Th}^{STN}$ & $2$ & $\omega_{D2}^{GPi}$ & $0$\tabularnewline
$\omega_{FS}^{D1}$ & $0.1$ & $\omega_{STN}^{GPe}$ & $0.005$ & $\omega_{D2}^{GPe}$ & $0.5$\tabularnewline
$\omega_{FS}^{D2}$ & $0.1$ & $\omega_{STN}^{GPi}$ & $0.006$ & $\lambda$ & $0.2$\tabularnewline
$\omega_{SC}^{Th}$ & $1$ & $\omega_{GPe}^{STN}$ & $0.003$ & $E_{D1}$ & $-0.1$\tabularnewline
$\omega_{TRN}^{Th}$ & $0.45$ & $\omega_{GPe}^{D1}$ & $0.6$ & $E_{D2}$ & $-0.1$\tabularnewline
$\omega_{BG}^{Th}$ & $0.2$ & $\omega_{GPe}^{D2}$ & $0.6$ & $E_{STN}$ & $0.3$\tabularnewline
$\omega_{Th}^{D1}$ & $2$ & $\omega_{GPe}^{FS}$ & $0.001$ & $E_{GPe}$ & $0.3$\tabularnewline
$\omega_{Th}^{D2}$ & $2$ & $\omega_{GPe}^{GPi}$ & $0.0002$ & $E_{GPi}$ & $0.3$\tabularnewline
$\omega_{Th}^{FS}$ & $0.03$ & $\omega_{D1}^{GPi}$ & $1$ & $E_{Th}$ & $0.1$\tabularnewline
\hline 
\end{tabular}
\caption{\label{tab_param_CBG}{\bf Parameters of the BG model}}
\end{table}

\begin{figure}[h]
\includegraphics[scale=0.37]{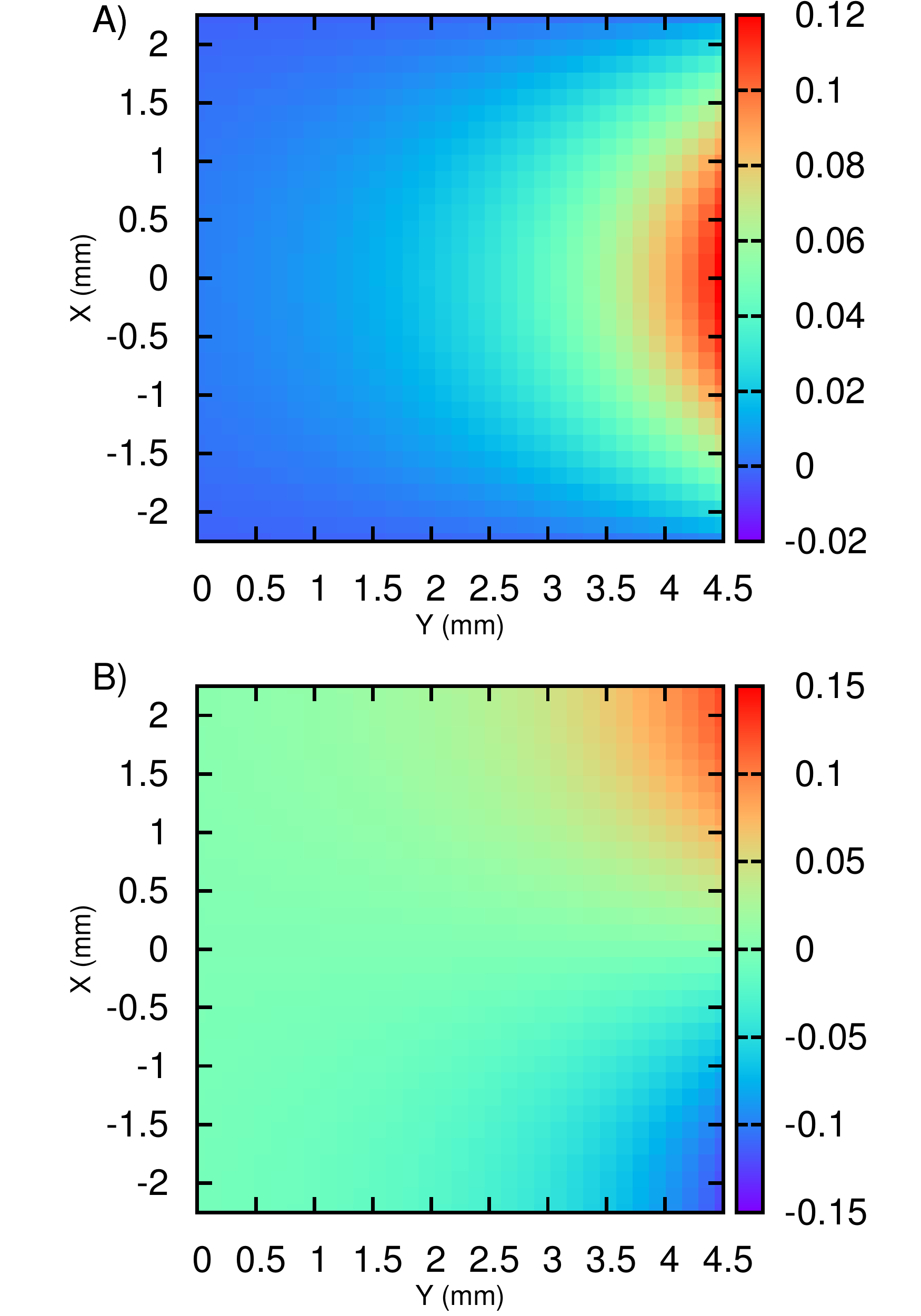}
\caption{\label{fig:annexe2_SC-SBG_weights}{\bf Maps of the connection weights from all the motor cells of the SC to the SBG}. A), projection from the SC to the horizontal SBG ; and B), projection from the SC to the vertical SBG. These maps were obtained according to Eq. (3) of \cite{Tabareau2007}, with the following monkey-specific parameters: $A=3\,^{\circ}$, $B_{X}=B_{Y}=1.5mm$, $a=1$, $b=0$.}
\end{figure}

\section*{References}

\bibliography{biblio_article.bib}

\end{document}